\pdfoutput=1
\documentclass[conference]{IEEEtran}
\usepackage[table,dvipsnames]{xcolor}
\IEEEoverridecommandlockouts

\usepackage{float} 
\usepackage{amsmath,amssymb,amsfonts}
\usepackage{algorithmic}
\usepackage{graphicx}
\usepackage{subfig}
\usepackage{bm}
\usepackage{enumitem}
\usepackage{tcolorbox}
\usepackage{wrapfig}
\usepackage{multirow}
\usepackage[utf8]{inputenc}
\usepackage{threeparttable}
\usepackage[numbers,sort&compress]{natbib}

\def\BibTeX{{\rm B\kern-.05em{\sc i\kern-.025em b}\kern-.08em
    T\kern-.1667em\lower.7ex\hbox{E}\kern-.125emX}}
    
\begin{document}
\newboolean{showcomments}
\setboolean{showcomments}{true}
\ifthenelse{\boolean{showcomments}}
{\newcommand{\mynote}[2]{
  \fbox{\bfseries\sffamily\scriptsize#1}
  {\small
  $\blacktriangleright$
  \textsf{{{\em #2}\bf}}
  $\blacktriangleleft$}}
}

\title{PEEK: Phishing Evolution Framework for Phishing Generation and Evolving Pattern Analysis using Large Language Models}

\author{\IEEEauthorblockN{1\textsuperscript{st} Fengchao Chen}
\IEEEauthorblockA{\textit{}
\textit{Monash University, CSIRO's Data61}\\
Melbourne, Australia \\
fengchao.chen@monash.edu}
\and
\IEEEauthorblockN{2\textsuperscript{nd} Tingmin Wu}
\IEEEauthorblockA{\textit{}
\textit{CSIRO's Data61}\\
Melbourne, Australia \\
Tina.Wu@data61.csiro.au}
\and
\IEEEauthorblockN{3\textsuperscript{rd} Van Nguyen}
\IEEEauthorblockA{\textit{}
\textit{Monash University, CSIRO's Data61}\\
Melbourne, Australia\\
Van.Nguyen1@monash.edu}
\and
\IEEEauthorblockN{4\textsuperscript{th} Shuo Wang}
\IEEEauthorblockA{\textit{}
\textit{Shanghai Jiao Tong University}\\
Shanghai, China \\
shuo.wang.edu@hotmail.com}
\and
\IEEEauthorblockN{5\textsuperscript{th} Alsharif Abuadbba}
\IEEEauthorblockA{\textit{}
\textit{CSIRO's Data61}\\
Melbourne, Australia \\
alsharif.abuadbba@gmail.com}
\and
\IEEEauthorblockN{6\textsuperscript{th} Carsten Rudolph}
\IEEEauthorblockA{\textit{}
\textit{Monash University}\\
Melbourne, Australia \\
Carsten.Rudolph@monash.edu}
}
\maketitle

\begin{abstract}
Phishing remains a pervasive cyber threat, as attackers craft deceptive emails to lure victims into revealing sensitive information. While Artificial Intelligence (AI), in particular, deep learning, has become a key component in defending against phishing attacks, these approaches face critical limitations. The scarcity of publicly available, diverse, and updated data, largely due to privacy concerns, constrains detection effectiveness. As phishing tactics evolve rapidly, models trained on limited, outdated data struggle to detect new, sophisticated deception strategies, leaving systems and people vulnerable to an ever-growing array of attacks. We propose the first \underline{P}hishing \underline{E}volution Fram\underline{E}wor\underline{K} (PEEK) for augmenting phishing email datasets with respect to quality and diversity, and analyzing changing phishing patterns for detection to adapt to updated phishing attacks. Specifically, we integrate large language models (LLMs) into the process of adversarial training to enhance the performance of the generated dataset and leverage persuasion principles in a recurrent framework to facilitate the understanding of changing phishing strategies. PEEK raises the proportion of usable phishing samples from 21.4\% to 84.8\%, surpassing existing works that rely on prompting and fine-tuning LLMs. The phishing datasets provided by PEEK, with evolving phishing patterns, outperform the other two available LLM-generated phishing email datasets in improving detection robustness. PEEK phishing boosts detectors' accuracy to over 88\% and reduces adversarial sensitivity by up to 70\%, still maintaining 70\% detection accuracy against adversarial attacks.
\end{abstract}


\section{Introduction}\label{chapter_introduction}
Email phishing is one of the most common attacks in phishing campaigns, which poses threats to individuals and organizations globally \cite{gupta2017fighting}. In the fourth quarter of 2024, losses attributed to reported phishing email attacks exceeded \$3.6 billion worldwide, nearly double the third quarter \cite{APWG_phish_report}, highlighting the significance of detecting phishing emails. Currently, deep learning (DL) \cite{salloum2022systematic,alhogail2021applying} and large language models (LLMs) \cite{koide2024chatspamdetector,chen2024survey} have been identified as the most powerful phishing detection approaches. Their performance is highly dependent on the characteristics of the available data, as these models identify phishing email campaigns based on the features learned from the training dataset. Ideally, the datasets for training detectors should be of high quality to be realistic, diverse across domains, and represent up-to-date phishing attacks to improve the robustness of detectors \cite{he2020towards}. 

Yet, challenges remain in collecting phishing emails with high quality and diversity, not least due to privacy concerns. Phishing emails that are reported to vendors are not shared publicly to protect users' private information from leaking. Additionally, phishing strategies change dynamically, while delays in the updating of phishing attacks result in detector training based on outdated datasets and a limited range of phishing subjects \cite{iwaya2023privacy,das2019sok}. To solve the limitations, researchers have made efforts to expand available phishing datasets by deploying DL models in conjunction with feature engineering (FE) \cite{salloum2021phishing} and data augmentation techniques \cite{regina2020text,shirazi2019adversarial,alkadri2022enhancing}. Despite promising results, FE-based methods still insufficiently capture phishing features \cite{das2019sok} and therefore achieve limited coverage of diverse phishing strategies. Concerning data augmentation, automatic methods such as adversarial training~\cite{tang2023data,xu2022adversarial} outperform rule-based approaches \cite{zhou2022rule} in reducing subjective intervention. Yet, the augmented datasets exhibit a diversity bias that hinders their effectiveness~\cite{chen2023empirical}.

Recently, LLMs have been utilized to generate phishing datasets due to their advanced generation capability. The IWSPA\_2023 phishing datasets \cite{mehdi2023adversarial} were provided through a fine-tuned GPT-2 that employs perturbation tools, aiming to defend against phishing threats. 
Nevertheless, the phishing emails provided by the method rely heavily on available datasets and with the risk of noisy outputs \cite{alomar2023data}, hindering the progress in phishing quality and diversity. Another phishing email dataset is synthesized by \textit{DeepAI}, an LLM-based generation tool, through prompts with phishing attempts and available Application Programming Interfaces (APIs) \cite{eze2024analysis}. However, \textit{DeepAI-phishing} lacks an assessment of phishing capability, which raises questions about whether realistic phishing patterns are preserved \cite{das2019sok}, potentially impairing the effectiveness of detectors due to inadequate feature learning.

Although LLMs have considerable potential for generating phishing emails \cite{zieni2023phishing}, systematic methodologies for evaluating LLM-generated phishing quality and enhancing diversity remain underdeveloped. Additionally, a subsequent challenge lies in understanding whether and how phishing patterns exploited by LLM-generated phishing diverge from those found in existing phishing, which is critical for comprehending unseen phishing threats. Furthermore, the question remains as to whether and how detectors can effectively improve robustness against changing phishing patterns and adapt to changing phishing tactics if training data does not reflect this dynamicity. 
To tackle the aforementioned issues,
we propose PEEK, an innovative Phishing Evolution Framework that leverages an adversarial training process to enhance the quality of synthetically generated phishing data, diversify phishing topics, and provide a long-term solution for interpreting changing phishing patterns.

To solve the challenge of limited phishing datasets, we employed and trained one of the state-of-the-art (SOTA) generative LLMs, Llama 3.1 8B, and integrated it within a generative adversarial network (GAN), which consists of a generator providing phishing content and a discriminator distinguishing between given and realistic phishing content. We leveraged the adversarial training process to enhance the generation capability and refine the quality of the provided phishing samples. The generated set is fed into an LLM-based analyzer for phishing validation, and we retain samples that are labeled as phishing. To comprehend new emerging patterns, we identify differences and cluster into groups to reveal distinctions in deceptive strategies from initial training datasets. Results (i.e., evolving phishing topic keywords) are incorporated into crafted chat prompts to strengthen diversity; these prompts are then fed back for further iterations, building PEEK as a recurrent framework that is executed iteratively. 
We ran two iterations of the PEEK framework, which enables large-scale phishing-sample generation and downstream analysis while minimising overfitting. The resulting PEEK phishing corpus and accompanying models will be released publicly upon the acceptance of the paper.

We evaluated the utility of PEEK phishing based on several criteria, including quality, diversity, and robustness. \textbf{Quality} assessment is based on two types of metrics: 1) To quantify the model's generation capability, LLM generation coherence (LLM-GC) and perplexity (PPL) are used, and scores of 0.83 and 6.95, respectively, indicate that the generator is effective in generating logically coherent and contextually reliable phishing samples. 2) To quantify the quality of LLM-generated phishing, the phishing authentication score (PAS) obtained from phishing analyzers is employed. A PAS analysis identifies 84.8\% realistic phishing mimicry. For \textbf{diversity}, we employed isolation forest score (IFS) and deceptive persuasion score (DPS) to identify anomalous phishing patterns and quantify persuasion strategies. The IFS revealed a statistically significant difference in features, and differences in DPS suggest that deception strategies in PEEK phishing differ from those in existing phishing techniques. For instance, PEEK phishing demonstrates varied vocabulary that reduces urgency, potentially decreasing recipients' alertness to the phishing intent. Regarding \textbf{robustness}, we evaluated PEEK phishing effectiveness by assessing the impact on detector robustness, using metrics such as accuracy, F1-score, and attack success rate (ASR). The results show a significant reduction in ASR, with detectors resisting up to 99.2\% while maintaining an F1-score above 0.88. Furthermore, our experiments in various phishing topics generation achieved a PAS analysis of 88.1\%, demonstrating the extensibility and adaptability of PEEK.

\noindent Ultimately, we provide the following key \textbf{contributions}:
\begin{itemize}[leftmargin=*]
\item We propose the Phishing Evolution Framework (PEEK), leveraging adversarial training to enhance and expand the utility of phishing emails. The iterative framework allows for generating and tracing the continuously changing tactics of LLM-generated phishing samples. 
\item We perform a comprehensive analysis with the framework based on seven real-world datasets, uncovering key characteristics of LLM-driven phishing generation.
\item Our framework is capable of generating phishing samples that exhibit both high quality and increased diversity, thereby improving the robustness of detectors against adversarial attacks.
\item We incorporate persuasion principles and linguistic analysis to deeply understand the difference between LLM-phishing and existing phishing datasets.
\end{itemize}

\section{Research Questions}\label{chapter_research_rqs}
\subsection{The problem statement}\label{chapter_problem_statement}
The increasing development of LLMs has intensified the economic losses caused by phishing attacks, while the limited availability and diversity of phishing datasets constrain the evolution of phishing defense technology. Additionally, the current extracted phishing features are unsuitable for explaining dynamically changing LLM-phishing strategies. Therefore, we propose a framework, PEEK, for phishing generation and deceptive features analysis. It aims to deliver high-quality, diverse phishing samples that continuously highlight the distinctions in social engineering strategies between PEEK phishing and existing datasets.

\subsection{Research Questions}\label{chapter_RQs}
\noindent We aim to answer the following questions:\\
\noindent \textbf{RQ1: How to leverage LLMs to enhance the quality and diversity of phishing emails based on available phishing datasets?} Enhancing the quality and diversity of phishing samples is crucial for improving the robustness and generalization of phishing detectors. Recently, motivated by the extensive general language training corpora, LLMs have been broadly applied to augment phishing datasets. Despite LLMs exhibiting impressive capability in generating contextually relevant text, concerns remain about whether the model faithfully captures key attack features from given training datasets. The absence of such features may undermine the effectiveness of generated phishing samples when used for detector training, thereby necessitating an assessment of the realism of synthetic phishing samples. We propose a novel pipeline that leverages adversarial principles to enhance the realism of LLM-phishing emails and employs chat-based prompt generation to diversify attack scenarios within ethical usage.

\noindent \textbf{RQ2: What are the differences in persuasive phishing strategies between existing and LLM-generated phishing datasets?} The extensive pre-training of LLMs enables them to potentially enrich phishing datasets with phishing texts that are both novel and domain-diverse, potentially creating persuasive strategies that diverge from strategies represented in existing datasets. These results can provide insights into changing attack strategies. Analyzing synthetic phishing data can benefit from recent work using LLM-driven phishing email detection for phishing intention detection \cite{heiding2023devising,bethany2024large}. However, these approaches provide limited explainability, which impedes the derivation of actionable insights on phishing strategies \cite{li2023survey}. Therefore, PEEK involves a linguistic tool to analyze the persuasive differences between existing and LLM-crafted phishing emails, aiming to enhance the interpretability of LLM-featured phishing strategies for developing proactive defense strategies.

\noindent \textbf{RQ3: Can iteratively extracted LLM-phishing patterns help detectors adapt to evolving phishing attacks?} Keeping phishing detectors up-to-date with changing phishing attack patterns is crucial for maintaining detector effectiveness in real-world deployment. Phishing attacks change dynamically, with phishers adapting their strategies to circumvent existing anti-phishing countermeasures. Thus, DL-based phishing detection based on fixed phishing patterns will always be disadvantaged against phishing strategies not yet sufficiently covered in current phishing classification~\cite{das2019sok,gryka2024detection}; Another challenge is the sensitivity of LLM-based phishing to perturbative attacks~\cite{de2025phishing}. In this work, we fill the gap of evolving attacks detection as much as possible by expanding quality phishing emails covering various attack patterns (i.e., persuasion principles and across topics). The supplementary attack patterns are then fed back into the framework, forming an iterative loop of ``generation-extraction'' to extend coverage to include targets and strategies previously not sufficiently covered.

\section{The Phishing Evolution Framework (PEEK)}\label{chapter_pen}
\begin{figure*}[t]
\centering
\includegraphics[width=1\linewidth, clip, trim=0.3cm 0.3cm 0.3cm 0.3cm ]{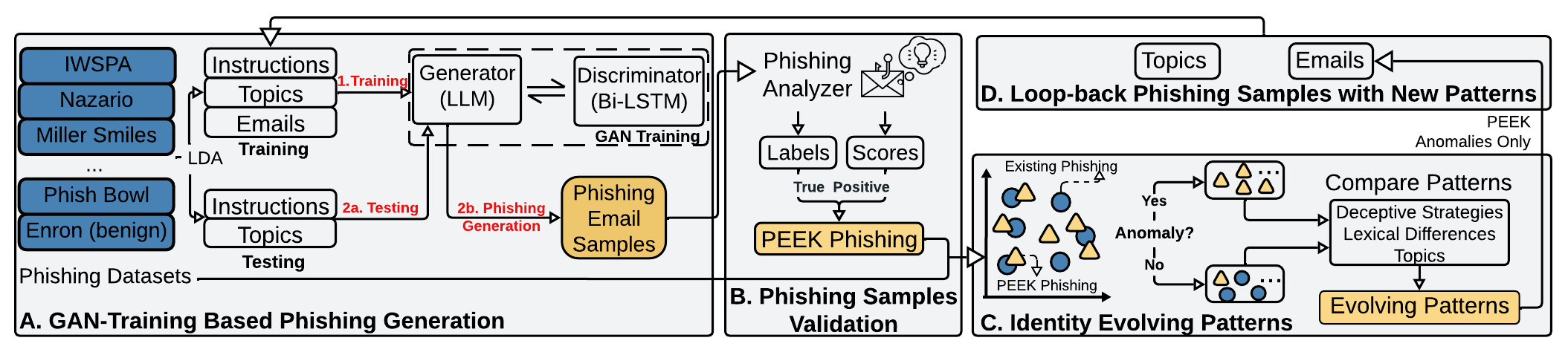}
\caption{Structure, phases and steps of the Phishing Evolution Framework (PEEK). 
The PEEK integrates a GAN model in Phase A to craft phishing emails, which are subsequently filtered in Phase B to isolate the best-quality phishing samples. In Phase C, the process incorporates persuasion principles alongside linguistic analysis to pinpoint emerging phishing tactics. The final phase, Phase D, utilizes these identified patterns to update phishing email samples in Phase A.}
\label{fig_pen}
\vspace{1mm}
\end{figure*}
Fig.~\ref{fig_pen} presents an overview of our proposed Phishing Evolution Framework (PEEK). 
The goals are to i) generate phishing samples with high quality and diversity; ii) understand how different LLMs leverage social engineering features deceiving users compared with existing phishing; iii) identify unique social engineering patterns from the evolution through a reasoning process; and iv) meet the need for a qualitative dataset and the insights into varied SE patterns to improve the current phishing detection mechanism.

The PEEK consists of four phases, starting with the use of existing phishing datasets and the GAN principle to guide the generator in producing new human-like phishing samples (Section \ref{chapter_pen_A}).
Next, we utilize a state-of-the-art phishing analyzer to assess the quality of generated phishing samples (Section \ref{chapter_pen_B}). We select the samples that are identified as phishing, referred to as PEEK phishing, which are subsequently prepared to identify new attacker strategies and changing characteristics. We specifically employ an anomaly-identifying algorithm to find differences with statistical significance, followed by unsupervised clustering and a linguistic tool for lexical level analysis to uncover evolving phishing patterns (Section \ref{chapter_pen_C}).
The extracted patterns are then fed back into the first phase of the framework, initiating a new iteration (Section \ref{chapter_pen_D}). Each iteration provides feedback to ensure that critical phishing characteristics are identified and analyzed. The pipeline is designed for generalization and can be updated as new LLMs are released, offering a long-term solution to the evolving phishing landscape. 
\subsection {\textbf{GAN-Training Based Phishing Generation}} \label{chapter_pen_A}
In the first stage of PEEK, real-world phishing emails are used to train the generator network (LLM-G) via the adversarial generative network (GAN) principle. The LLM-G and a discriminator (D) compete and refine each other, ultimately enhancing the generator's ability to produce more realistic phishing samples. The complete training procedure is outlined in Appendix~\ref{appendix:object_function}. This process, termed GAN-based training, is designed to generate content with high quality and diversity, allowing us to observe how output evolves after fine-tuning and revealing shifts in social engineering tactics. 

\begin{figure}[ht]
\centering
\includegraphics[width=1\linewidth]{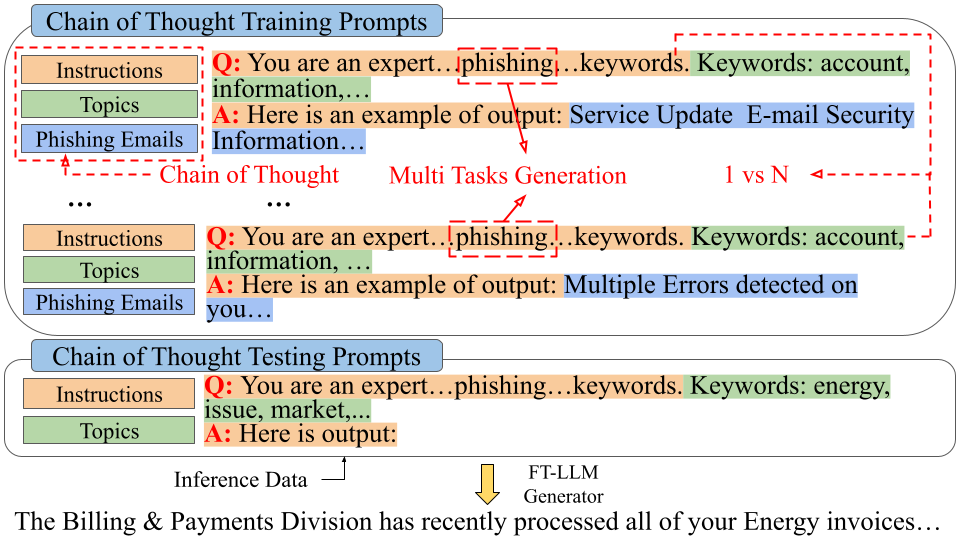}
\caption{Chain of Thought Designed Prompts. Training prompts consist of \textit{Instruction, Topics,} and \textit{Existing Data.} \textit{Instructions} frame the model as a cybersecurity expert, \textit{Topics} guide generation, and \textit{Existing Data} provides output example data. During testing, the generator gives PEEK phishing samples given by the prompt that excluded \textit{Existing Data}.
}
\label{fig_llama_prompt_example}
\end{figure}

\noindent \textbf{Generator Network.} We deploy a Llama 3.1 8B model (referred to as Llama 3.1 in the paper) with a custom-designed prompt (refer to Fig.~\ref{fig_llama_prompt_example}), as the generator of GAN. Compared with other LLMs (e.g., Claude-3~\cite{claudeai}, Gemini~\cite{team2023gemini}), Llama~3.1 stands out as an open-source model incorporating the latest text generation technology advancements. This makes it particularly valuable for generating phishing emails that reflect current trends, ensuring its relevance for research and deceptive tactics development. To fine-tune Llama 3.1 efficiently, we apply the Low-Rank Adaptation (LoRA) technique \cite{hu2021lora}, which is well-suited for large pre-trained models with limited computational resources.

The LLM generator leverages latent representations to simulate features from real text data and provides samples that closely mimic real-world phishing content. Through adversarial training, the discriminator differentiates between real and generated content while providing feedback to the generator. We conducted an ablation study to compare the performance of different generators (e.g., GPT-2) with Llama 3.1 in phishing generation with that of different states (pre-trained and fine-tuned). The methodology, configuration, and results can be found in Appendix \ref{appendix:chapter_prompts_ablation_study}.


\textbf{Discriminator Network}. For this task, we employ a Bidirectional Long Short-Term Memory (Bi-LSTM) network \cite{jang2020bi} to assess the realism of the generated samples from the generator. Bi-LSTM is an enhanced version of the recurrent neural network (RNN), incorporating input, forget, and output gates to effectively model long-term dependencies in sequences. Unlike standard LSTMs, Bi-LSTMs process input sequences in both forward and backward directions, allowing the model to capture contextual information from both preceding and succeeding tokens. This bidirectional context significantly enhances performance in text classification tasks, particularly when the meaning of a token depends on its surrounding context. Although large language models (LLMs) have shown strong performance in text classification \cite{uddin2024explainable}, employing LLMs as a discriminator poses a risk of classification bias induced by prompting-specific preference, and their substantial parameter size introduces significant computational overhead in adversarial training. Thus, to balance execution efficiency and detection accuracy, we selected a Bi-LSTM as the discriminator.

\noindent \textbf{Prompts for Generation Guidance.}
To effectively guide the generation during GAN training, we avoid directly inputting the filtered phishing samples into the generation models. Instead, we designed a structured chat prompt template inspired by the Chain of Thought (CoT \cite{wei2022chain}). This template consists of three key sections: Instructions, Topics, and Phishing Emails~(Fig.~\ref{fig_llama_prompt_example}). Each section is designed to incrementally guide the content generation process as follows:
\begin{itemize}[leftmargin=*]
\item Instructions: This section sets the context by defining the model as a cybersecurity expert, helping to minimize hallucinations and focus the model’s responses on specific tasks. Notably, including an email label here is essential for specifying the type of email (e.g., phishing or legitimate). Without this label, the model might still generate content resembling legitimate emails, even if phishing-related keywords exist.
\item Topics: A list of keywords serves as guiding constraints, directing the model’s creativity while ensuring coherence with the desired themes or topics. This approach enhances topic consistency, enabling the generation of diverse content that remains relevant to the phishing theme.
\item Phishing Emails (Existing Data): This section provides examples of both phishing and legitimate emails, serving as output templates. These examples help clarify the expected response format, guiding the model to interpret the given instructions and generate plausible content.
\end{itemize}
 
The design incorporates a one-to-many relationship between the output templates and topics, preventing limitations in the diversity of the generated PEEK content. As outlined in Section \ref{chapter_pen_A}, the generator is then fine-tuned using the comprehensive prompt in GAN, triggering a new iteration of the process. In this step, we obtain an initial generated dataset for the framework, labeled as PEEK-generated phishing samples.

\subsection{\textbf{Phishing Samples Validation}}\label{chapter_pen_B}
The goal of the second stage is to validate the realism of the generated results (RQ1). This approach provides a set of high-quality, human-like phishing samples for further analysis. We deploy ChatGPT as an analyzer to validate and retain PEEK-generated phishing samples that are classified as ``phishing'' based on the prompt designed by Koide \cite{koide2024chatspamdetector}. The prompt performs a thorough analysis of phishing attempts and quantifies phishing determination. Phishing samples after this step are labeled as PEEK phishing. The analyzer undergoes a reliability test using cross-validation and is assessed based on the results of F1-score and F2-score, which ensure that identified samples are labeled as phishing with high confidence. We refer to the final output as PEEK phishing emails. Unlike standard SOTA methods (e.g., Gmail and Outlook filters), our models focus on deeper contextual phishing detection, particularly targeting the email body discovery. 

\subsection{\textbf{Identity Evolving Patterns}}\label{chapter_pen_C}
To understand the diverse phishing strategies that emerge in PEEK phishing (RQ2), we start by applying an anomaly identification algorithm to identify samples that differ from both existing and PEEK phishing emails. We refer to the PEEK phishing emails that are distributed within the majority phishing set but labeled as outliers as PEEK anomalies. These anomalies are then clustered into several interpretable categories for persuasive patterns analysis. Specifically, we used the isolation forest score (IFS) for anomaly identification. IFS is commonly applied in anomaly phishing detection and is generally robust to data noise \cite{liu2008isolation}. Latent Dirichlet Allocation (LDA) is applied for text clustering, as it can automatically discover topics without requiring predefined categories~\cite{blei2003latent}. The optimal number of cluster topics is determined by the text coherence scores, which range from 0 to 1, with higher scores indicating stronger coherence. 

We then delve into the deceptive characteristics of each anomaly cluster, focusing on Cialdini's persuasion principles~\cite{cialdini2001science}: \textit{Authority, Reciprocity, Scarcity, Liking, and Social Proof}. These principles are commonly leveraged by attackers to shape human behavior and influence decision-making. For instance, attackers might employ these tactics to create a sense of urgency and induce panic, thereby misleading victims into making irrational decisions. One of the goals of this study is to compare the persuasive tactics employed in traditional phishing with those found in LLM-generated phishing content. 

To annotate persuasion principles, we leverage LLOOM, an integrated tool that distills input data, clusters information, and outputs summarized concepts \cite{lam2024concept}. By providing the conceptual definitions of persuasion principles, LLooM performs inverse annotation to identify the principles within PEEK anomalies. To gain deeper insights into specific principles (e.g., differences in lexical usage), we employ Linguistic Inquiry and Word Count 2022 (LIWC-22) \cite{boyd2022development}. LIWC-22 quantifies linguistic features using a pre-compiled dictionary that maps words to psychological insights, encompassing emotional, social, cognitive, and authority-based language categories \cite{boyd2022development}. These insights align with psychologically meaningful word groups, each corresponding to one or more persuasion principles. For instance, the \textit{Social Proof} principle reflects human tendencies toward conformity and peer influence, often indicated by terms like \textit{``everyone''} and \textit{``most people''}.

By leveraging these tools, we aim to 1) uncover deceptive characteristics through psychological and semantic analysis and 2) illustrate the evolution of phishing techniques, including variations in the use of persuasion principles, word choice, and emotional expression.

\subsection{\textbf{Loop-back Phishing Samples with New Patterns}}\label{chapter_pen_D} Our approach creates a recurrent iterative framework by reintegrating high-quality PEEK phishing samples into the training datasets. In this phase, we insert extracted features from Section \ref{chapter_pen_C} (i.e., topic keywords, labels, and email content) into the chat prompt used during training, starting a new cycle of training, generation, and analysis. This iterative loop enables the continuous refinement of the phishing sample dataset, integrating evolving patterns and themes while maintaining high-quality, diverse content generation that aligns with current phishing tactics.


\section{Evaluation Criteria}\label{chapter_metrics_evaluation_metrics}
The goal of PEEK is to generate diverse phishing datasets that closely mimic human-like phishing attempts.  To systematically evaluate the effectiveness of generated phishing emails, we introduce three primary evaluation criteria: \textit{Quality}, \textit{Diversity}, and \textit{Robustness}.

\noindent \textbf{Quality} assesses how closely the generated phishing emails resemble realistic phishing email campaigns. We evaluate quality from two aspects: the generator's performance and the realism validation of generated emails. To quantify the quality, we introduce the following metrics:
\begin{itemize}[leftmargin=*]
    \item \textbf{LLM Generation Coherence (LLM-GC)}: The metric measures the generative model's ability to produce contextually relevant outputs based solely on provided inputs (e.g., instructions, labels, keywords, and email prompts) without relying on external domain knowledge. The LLM-GC yields a coherence score ranging between 0 to 1 \cite{mimno2011optimizing}, where scores closer to 1 reflect higher contextual consistency and relevance.
   
    \item \textbf{Perplexity (PPL)}: This metric quantifies the generative model's confidence in predicting subsequences in a given sequence, with values ranging from 1 upwards  \cite{brown1992class}. Lower perplexity values (close to 1) indicate content closely mirroring the input data but potentially risking ease of classification, whereas higher perplexity suggest increased randomness in the generated content. Typically, perplexity scores between 1 and 10 strike a balance between authenticity and variety.
    
    \item \textbf{Phishing Authentication Score (PAS)}: PAS evaluates the realism and authenticity of generated phishing content based on content structure, contextual relevance, and psychological manipulation tactics common in human-crafted phishing emails. PAS is determined through a ChatGPT analyzer employing targeted prompts to classify samples, assign phishing authenticity scores, and provide explanatory rationales \cite{koide2024chatspamdetector}. Scores range from 0 to 10; higher scores indicate high-quality, authentic phishing imitations, and scores below 5 indicate poor realism or non-phishing characteristics. The ChatGPT analyzer's reliability was validated using 5-fold cross-validation tests, achieving F1 and F2 scores of 0.95 and 0.97, respectively. Detailed validation processes are provided in Appendix~\ref{appendix:analyzer_CV}.
\end{itemize}

\noindent \textbf{Diversity} evaluates the novelty and variety of PEEK-generated phishing emails compared to existing datasets, focusing on thematic variance, deceptive strategies, and uniqueness of attack vectors. 
Metrics include:
\begin{itemize}[leftmargin=*]
    \item \textbf{Isolation Forest Score (IFS)}: This metric identifies novel or outlier phishing emails based on negative isolation scores, derived by isolating content through randomly selected feature splits \cite{liu2008isolation}. Text content is numerically represented using Term Frequency-Inverse Document Frequency  (TF-IDF) \cite{wu2008interpreting}, highlighting distinctive linguistic features across datasets.
    \item \textbf{Deceptive Persuasion Score (DPS)}: DPS quantifies the semantic alignment of generated phishing emails with predefined persuasive principles (e.g., authority, urgency). 
    It ranges between 0 and 1, computed via LLooM, a conceptual summarization tool \cite{lam2024concept}. 
    Higher scores indicate stronger semantic relevance, signifying explicit or implicit persuasive intent. Samples lacking persuasive principles are excluded from subsequent training.  
\end{itemize}
\noindent \textbf{Robustness} assesses the effectiveness of generated phishing content in enhancing the resilience and adaptability of phishing detection models against evasion attacks. Robustness is measured by analyzing detection performance improvements after fine-tuning on PEEK-generated phishing emails compared to two benchmark datasets  
(IWSPA\_2023 \cite{mehdi2023adversarial} and DeepAI-phishing \cite{eze2024analysis}). 
Metrics include: 
\begin{itemize}[leftmargin=*]   
    \item \textbf{Attack Success Rate (ASR)}: ASR~\eqref{eq:asr} represents the proportion of adversarial phishing emails successfully evading detection \cite{goodfellow2014explaining}. A higher ASR indicates greater model vulnerability and reduced detection robustness. 
    \begin{align}
        \text{Attack Success Rate (ASR)} = \frac{\text{Successful~Evasions}}{\text{Total~Evasions}}
        \label{eq:asr}
    \end{align}
\end{itemize}
Detection performance pre- and post-fine-tuning on the datasets is evaluated using \textit{precision}, \textit{recall}, \textit{accuracy}, and \textit{F1-score} \cite{allen1955machine}.

\section{Implementation}\label{chapter_implementation}
\subsection{Datasets}\label{chapter_implementation_datasets}
This study uses phishing email datasets from IWSPA-AP~\cite{IWSPA}, Nazario \cite{Nazario}, Miller Smile \cite{Miller_Smiles}, Phishing Bowl \cite{phish_bowl}, Nigerian Fraud \cite{Nigerian_Fraud}, and the Cambridge dataset\footnote{private dataset}, along with benign emails primarily sourced from Enron, supplemented by IWSPA-AP. The dataset comprises approximately 70,000 phishing emails and is diverse in structure. To ensure consistency, all datasets were preprocessed to extract and format email content in a standardized manner.

The open-source phishing emails pose several challenges: 1) duplicate entries across datasets; 2) structural and lexical redundancy; and 3) incomplete content, such as extremely short emails. To address the issues, we eliminate duplicates and perform similarity analysis to remove highly similar samples. Email lengths were standardized to meet the requirement with the pre-trained language model, with a maximum of 512 tokens and a minimum of 64 tokens to ensure sample quality. The final dataset comprises 13,800 processed phishing emails (see Appendix \ref{appendix:dataset_info}), with 80\% allocated for fine-tuning and 20\% reserved for initial inference.

\subsection{Model Configuration}\label{}
In each phase of PEEK, distinct LLMs are utilized for specific tasks. Llama 3.1 is employed to generate phishing emails in the generation stage, enhanced with LoRA for improved computational efficiency. 
Pre-trained ChatGPT, augmented with prompt engineering, is used for phishing verification. For assessing the practicality, we employed two types of detectors: 1) ALBERT, RoBERTa, and SQ \cite{mehdi2023adversarial}, which distinguish human-written legitimate from phishing emails; and 2) DistilBERT, RoBERTa-LLMs \cite{jamal2024improved} and T5-Encoder~\cite{bethany2024large}, which differentiate human-crafted phishing emails from LLM-generated. Perturbation tools were utilized to simulate real-world phishing detection and evasion attacks. Except for the ChatGPT analyzer, the relevant models were fine-tuned on an NVIDIA H100 GPU (93.25 GB VRAM), with detailed parameter settings provided in Appendix~\ref{appendix:model_config}.

\section{Experimental Results}\label{chapter_results}
This section addresses RQ1 by evaluating the quality and diversity of LLM-generated phishing emails using the metrics detailed in Section \ref{chapter_metrics_evaluation_metrics}. 
\subsection{RQ1: LLM-generated phishing Quality and Diversity.}\label{chapter_RQ1}

\subsubsection{\textbf{Quality Assessment}}

\begin{figure*}[t]
  \centering
  \begin{minipage}[t]{0.32\textwidth}
    \centering
    \includegraphics[width=\linewidth,height= 4.8cm,clip,trim=0.3cm 0.5cm 0.3cm 0.3cm]{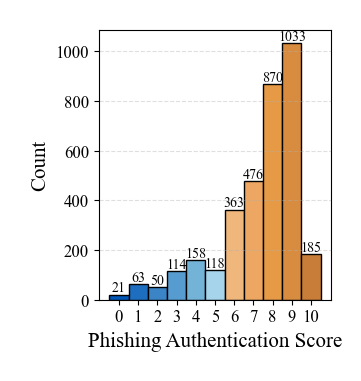}
    \caption{Phishing Authentication Score. Higher scores indicate realistic phishing generations, and scores below 5 denote low phishing authenticity.}
    \label{fig_phishing_score_distributions}
  \end{minipage}\hfill
  \begin{minipage}[t]{0.32\textwidth}
    \centering
    \includegraphics[width=\linewidth]{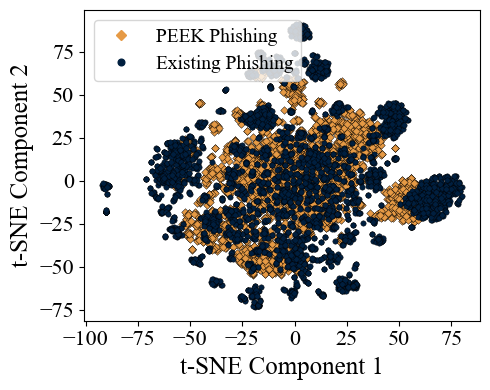}
    \caption{Mapped distribution of Existing and PEEK phishing emails. Overlap indicates semantic similarity while distinctiveness represents variants.}
    \label{fig_distribution_differences}
  \end{minipage}\hfill
  \begin{minipage}[t]{0.32\textwidth}
    \centering
    \includegraphics[width=\linewidth,clip,trim=0 0 0 0]{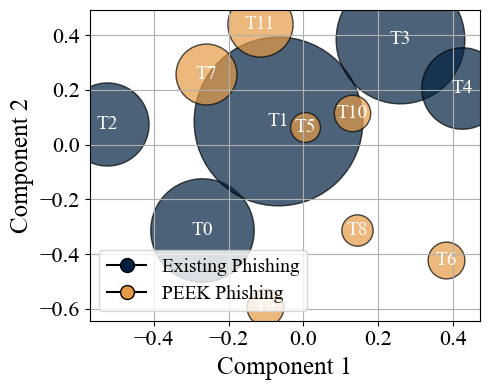}
    \caption{Latent Phishing Topics Distribution. Topic circle size denotes weight, distance shows similarity. Dispersion reflects phishing diversity.}
    \label{fig_expanded_topic}
  \end{minipage}
\end{figure*}

To enhance the realism of generated phishing samples, we leveraged the adversarial training principle, incorporating a discriminator to iteratively refine the outputs to be realistic. Table~\ref{tab:generators_selection_filters_results} indicates that PEEK generates 63.4\% and 4.2\% more validated phishing samples compared with pre-trained and fine-tuned Llama 3.1 models, respectively, highlighting the benefit of the GAN-based training. We further assess how closely the generated emails resemble authentic phishing emails, with a focus on generative performance and realism metrics. The LLM Generation Coherence (LLM-GC) and Perplexity (PPL) metrics quantify the generative model’s effectiveness. PEEK achieved an LLM-GC score of 0.83, indicating a strong adherence to input prompts and alignment with intended phishing goals. A PPL of 6.95 demonstrates enhanced generation capability of PEEK, with output samples that are fluent and closely aligned with classical phishing language patterns (the overlap shown in Fig.~\ref{fig_distribution_differences}). In summary, these results demonstrate PEEK's proficiency in generating logically coherent and contextually aligned quality phishing content.

\begin{figure}[th]
  \centering
  \subfloat[PEEK phishing Email with analyzer quantify score of 6.]{%
    \includegraphics[width=0.95\columnwidth]{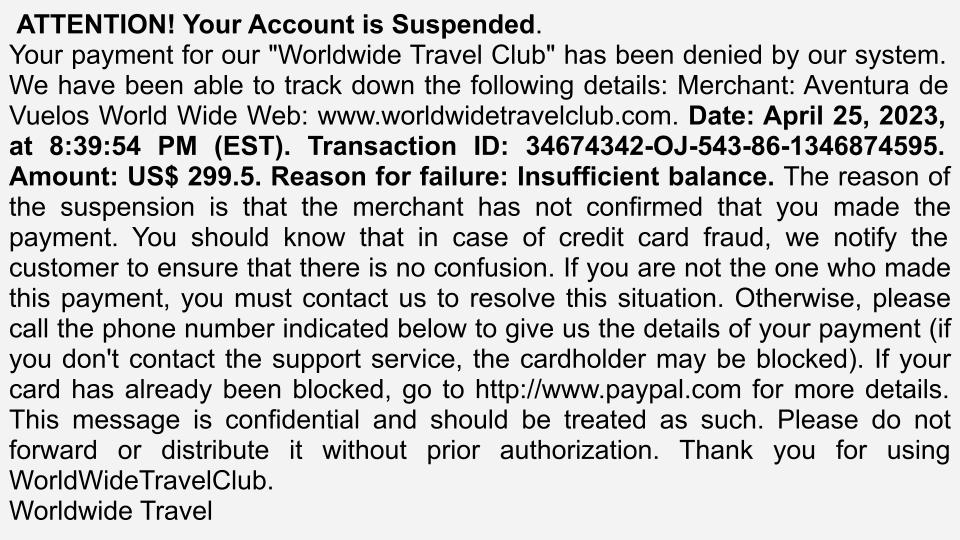}%
    \label{fig_pen_phishing}%
  }\\
  \subfloat[PEEK non-phishing Email with analyzer quantify score of 2.]{%
    \includegraphics[width=0.95\columnwidth,clip,trim=0cm 10cm 0cm 0cm]{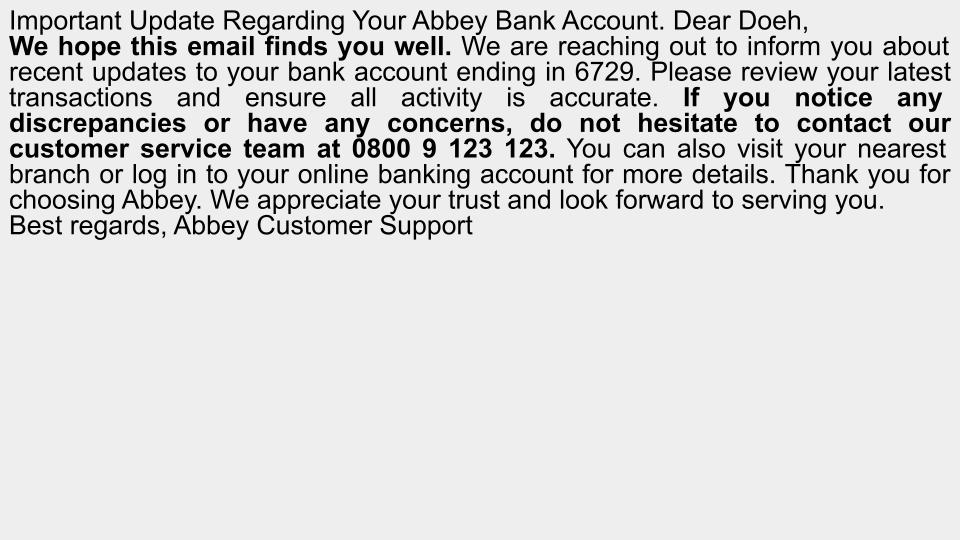}%
    \label{fig_pen_non_phishing}%
  }
  \caption{Examples of PEEK-generated Email Sample}
\end{figure}

The Phishing Authentication Score (PAS) further validates the realism of PEEK phishing samples. As shown in Fig.~\ref{fig_phishing_score_distributions}, overall, 84.8\% of PEEK-generated phishing samples received PAS scores of 6 or higher, indicating realistic phishing attempts. Specifically, 71.3\% obtained scores of 8 or above, characterized by common phishing triggers such as urgency cues (e.g., \textit{``alert notifications''} or immediate action demands). These PEEK phishing samples exhibit strong alignment with existing phishing deceptive vectors. The remaining 13.5\% of PEEK phishing samples received a score of 6 or 7, showing uncertainty in phishing classification. Further analysis revealed that this subset features precise contextual elements (e.g., transaction IDs and times that are highlighted in Fig.~\ref{fig_pen_phishing}), alongside common phishing indicators such as urgency and scarcity. The details enhance the perceived authenticity and illustrate variations in phishing strategies, supporting further investigation into LLM-driven differences in phishing email attacks. Email samples scoring below 5 lack explicit phishing attempts, as exemplified in Fig.~\ref{fig_pen_non_phishing}.

\begin{tcolorbox}[colback=gray!20,boxsep=1mm, left=0.125mm, right=0.125mm, top=0.125mm, bottom=0.125mm]
\textbf{Takeaway 1}: PEEK demonstrates strong performance in generating high-quality phishing emails. Metrics (LLM-GC=0.83, PPL=6.95) indicate PEEK is effective in generating logically coherent and contextually reliable phishing samples. PAS analysis confirms over 84\% of PEEK-generated phishing emails align closely with human-crafted phishing content.
\end{tcolorbox}

\subsubsection{\textbf{\textit{Diversity in Expanded Phishing Topics}}}
To identify divergences in phishing topics, we employed Latent Dirichlet Allocation (LDA) clustering to uncover latent thematic differences between PEEK phishing and existing phishing emails. As illustrated in Fig.~\ref{fig_expanded_topic}, the analysis indicates PEEK significantly expands the scope of phishing topics, facilitating richer and more varied phishing expressions.

Existing phishing email topics predominantly cover two main themes: 1) banking services related to account security, including online service feedback (T0), verification processes (T1), password updates (T2), and restricted account access (T4); and 2) business-oriented communications and extending company-related interactions (T3). While the majority of the PEEK phishing emails align closely with these established topics, LDA analysis reveals three additional emerging topics unique to PEEK. Specifically, topic T6 shifts away from typical banking themes toward shipment and delivery issues concerning \textit{contact address updates}. These emails frequently employ misleading claims about incorrect addresses to facilitate direct interactions, such as prompting recipients to engage via phone calls (example in Fig.~\ref{fig_diversity_example1}). Furthermore, topics T8 and T9 introduce additional diversity by addressing \textit{bank account login failures} and \textit{university online service interactions}, respectively. These findings demonstrate that PEEK can not only replicate existing phishing domains but also be triggered to diversify and innovate new phishing topics. The expanded phishing topics are given by open-ended keywords in crafted chat prompts, thus enabling flexible topic generation and enriching the simulated attacks landscape to enhance detector robustness against unseen threats.


\begin{tcolorbox}[colback=gray!20,boxsep=1mm, left=0.125mm, right=0.125mm, top=0.125mm, bottom=0.125mm]
\textbf{Takeaway 2}: LDA analysis demonstrates PEEK's capability to generate phishing topics that both align with existing phishing themes and expand into novel domains.
\end{tcolorbox}

\subsection{RQ2: LLM-driven Phishing Patterns that Differ from Existing Phishing Datasets}\label{chapter_RQ2}
\begin{figure}[th]  
    \centering
    \includegraphics[width=0.8\linewidth, clip, trim=0cm 0cm 0cm 0cm]{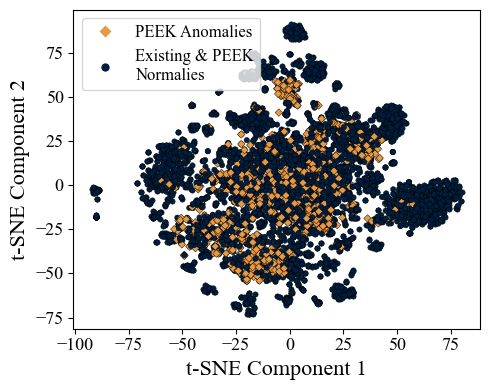}
    \caption{Identified PEEK Anomalies. Though PEEK anomalies appear spatially similar to the general phishing distribution, they exhibit significant lexical and persuasive distinctions, justifying focused analysis.}
    \label{fig_iso_anomalies_normalies}
\end{figure}

To address RQ2, which investigates whether and how LLM-generated phishing content differs from traditional phishing strategies in manipulating user perception, we applied the isolation forest (IFS) algorithm to identify subtle lexical outliers within a mixed dataset of existing phishing emails and LLM-generated phishing from the PEEK dataset. The analysis identified 945 PEEK samples as statistically distinct, hereafter referred to as PEEK anomalies ($PEEK_{anml}$)  (Fig.~\ref{fig_iso_anomalies_normalies}).  

\begin{figure*}[th]  
    \centering
    \includegraphics[width=\linewidth, clip, trim=0.3cm 0.5cm 0.3cm 0.3cm ]{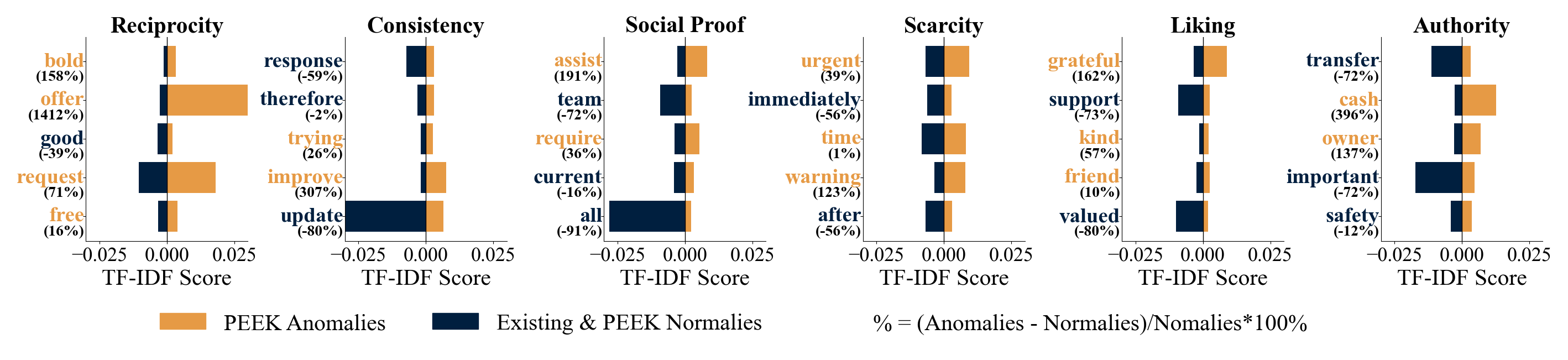}
    \caption{Linguistic Feature Differences Aligned with Persuasion Principles. Statistically significant features illustrate how specific persuasive strategies vary across normal and LLM-generated phishing.}
    \label{fig_iso_split_features}
\end{figure*}

To validate the lexical distinctiveness of these anomalies, we conducted a Mann-Whitney U test on IFS-selected features. Results confirmed significant differences between PEEK anomalies and normal samples (Fig.~\ref{fig_iso_split_features}), with lexical distinctions aligning with varied persuasive intentions. These findings suggest that LLMs adopt alternative manipulation strategies that diverge from previously observed phishing tactics. 

\begin{figure}[ht]
\centering
\includegraphics[width=1\linewidth, clip, trim = 0 0 0 0]{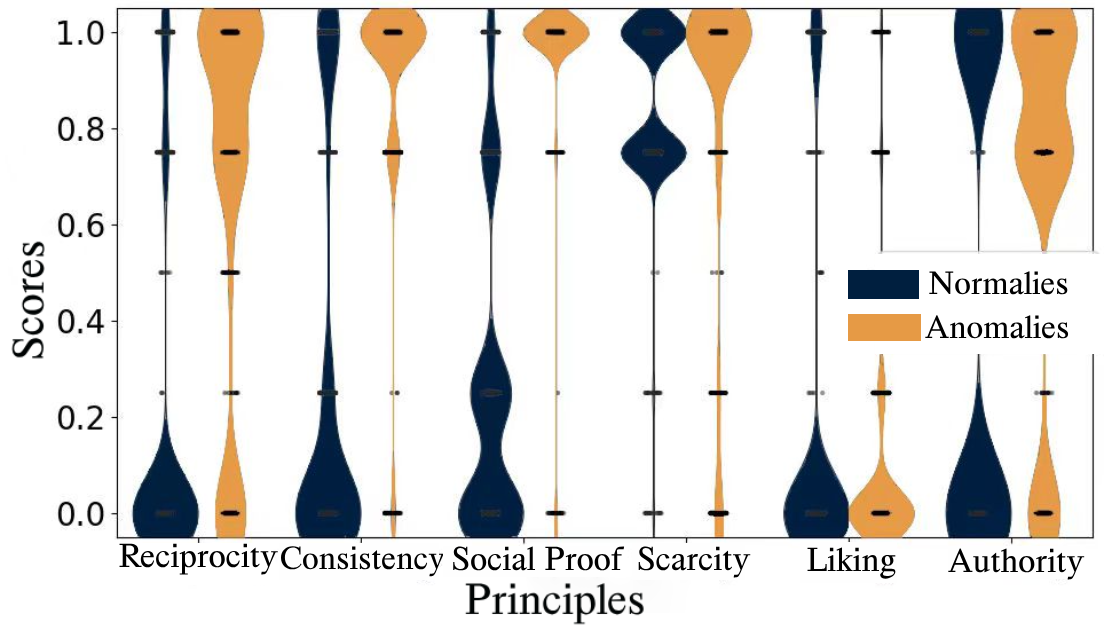}
\caption{Deceptive Persuasion Score (DPS) for Normal vs. PEEK Anomaly Phishing Emails. Higher scores indicate greater use of persuasive principles. PEEK anomalies display broader and more frequent use of persuasive strategies.} 
\label{fig_general_principles_difference}
\end{figure}

To further analyze the persuasion complexity of PEEK anomalies, we computed the Deceptive Persuasion Score (DPS), which estimates the number and intensity of persuasive principles within each sample. As shown in Fig.~\ref{fig_general_principles_difference}, PEEK anomalies consistently employ a wider array of principles with greater frequency. Notably, 2.13\% of PEEK anomalies contain all six persuasion principles, and 40.9\% include five. In contrast, only 0.78\% of traditional phishing samples use all six persuasion principles, and 17.11\% employ five. This divergence is reflected in the usage patterns of key lexical elements. For example, the phrase \textit{``offer cooperation''}, aligned with the Reciprocity principle, appears 14.12 times more frequently in PEEK anomalies than in traditional samples. 
These results highlight the increased rhetorical complexity in LLM-generated content, underscoring the need to reevaluate detection methods.

\begin{table*}
    \centering
    \caption{Common Words and Contextual Differences between Normal and PEEK Anomalies Phishing Emails. The column \textit{Principles} lists the first three persuasion principles identified, along with relevant LIWC categories. The column \textit{N-Matching} shows the most frequent words for normal phishing, and ``\textit{P-Matching}'' for PEEK anomalies. \textit{N-/P-Context} provides n-gram word differences before and after the matching words, highlighting contextual differences.}
    \begin{tabular}{llllll}
    \hline
     \textbf{Principles} & \textbf{LIWC Expression} & \textbf{N-Matching} & \textbf{N-Context} & \textbf{P-Matching} & \textbf{P-Context}\\
         \hline
         & \textit{power,} & royal & [account verification, Scotland bank] & client & [corruption of, data copyright]\\
Authority& \textit{work,} & authorized & [banking detail, bank electronic] & department & [clients service, part of group] \\
         & \textit{social referents} & official & [Lloyds online, notification Natwest] & service & [customer online, team that]\\
        \hline
         & \textit{time, certitude,} & urgent & [await your, mail please reply] & before & [obtain your permission, share] \\
Scarcity & \textit{need, lack,} & immediately & [hear from you, and provide your] & continue & [this process, until you reach] \\
         & \textit{risk} & must & [organize records, confirm account] & time & [update at any, is subject to] \\
        \hline
         & \textit{affiliation} & apologize & [bank account, for inconvenience] & instructions & [follow provided, to complete] \\
Reciprocity & \textit{socia l reference} & appreciates & [will be highly, thank you in] & support & [contact online, further help] \\
         & \textit{reward} & kin & [the next, to the account] & offer & [services we, to begin with]\\
        \hline
        & \textit{cognition,memory,} & have to & [program you will, re-confirm] & may & [contract, periodically send] \\
Consistency& \textit{achieve,} & make sure & [conditions, further actions need] & wish & [any point, to receive this] \\
         & \textit{focus future} & planned & [account is, assist reply this] & ready to & [statements, help responding] \\
         \hline
    \end{tabular}
    \label{tab:principle_frequent_words}
\end{table*}

\begin{figure}[ht]
\centering
\includegraphics[width=1\linewidth]{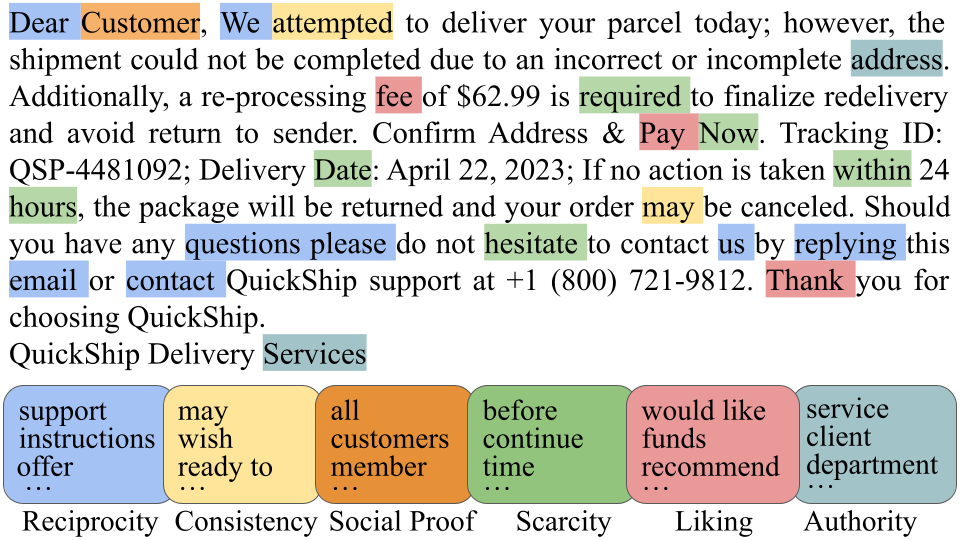}
\caption{Example of PEEK-generated phishing email with persuasive principle tags, using LIWC-22’s Color Coding. Highlighted terms correspond to persuasion principles.}
\label{fig_diversity_example1}
\vspace{1mm}
\end{figure}

Across the top persuasion strategies, \textit{Authority}, \textit{Reciprocity}, and \textit{Scarcity}, PEEK anomalies employ more concrete and contextualized language. 
For example, in \textit{Authority} principle, referencing credible institutions to build trust revealed a notable difference between the two datasets in how authority is conveyed ($median_{nml} = 1.2$, $median_{anml} = 0.86$). Overall, $PEEK_{anml}$ includes more concrete details that strengthen its persuasive impact. As shown in Table \ref{tab:principle_frequent_words}, the top three frequent \textit{Authority} related words differ markedly from those in normal phishing emails. PEEK anomalies often include specific references to prominent companies (e.g., HSBC, PayPal, Microsoft). It also extends to detailed institution markers, such as \textit{``xxx Client Center''}, and \textit{``xxx Service''} (example in Fig.~\ref{fig_diversity_example1}). In contrast, normal phishing emails tend to rely on broader, descriptive language, using terms like \textit{``royal''} and \textit{``authorized''} to imply authority.

The increased detail in PEEK anomalies is also evident in the use of the \textit{Reciprocity} principle ($median_{nml} = 0.86$, $median_{anml} = 1.24$). Terms like \textit{``instruction''} often include specific steps or commands, and \textit{``support''} typically specifies the scope of assistance provided. This linguistic specificity requires the use of logical connectors (e.g., \textit{``cause''}, \textit{``insight''}) to maintain coherence, which increases cognitive complexity and suggests deeper user engagement \cite{tausczik2010psychological}. In contrast, normal phishing emails rely on generic phrases such as \textit{``apology for''}, \textit{``appreciate your cooperation''} with minimal elaboration. 
This specificity introduces more cognitive load, increasing user engagement and perceived authenticity. \cite{xiao2019changing}. 

\begin{tcolorbox}[colback=gray!20,boxsep=1mm, left=0.125mm, right=0.125mm, top=0.125mm, bottom=0.125mm]
\textbf{Takeaway 3}: Over 42\% of PEEK anomalies incorporate five or more persuasive principles. Their lexical specificity and cognitive complexity significantly enhance their deceptive strength. 
\end{tcolorbox}

In contrast to traditional phishing’s overt urgency, PEEK anomalies demonstrate a strategic downplaying of \textit{Scarcity} signals 
($median_{nml} = 0.86$, $median_{anml} = 0.63$).
Urgency-related words such as \textit{``urgent''}, \textit{``immediately''}, and \textit{``must''}, or phrases like \textit{``limited time''} and \textit{``offer ends soon''} are common in phishing. 
However, recent studies suggest that users have become more adept at recognizing these urgency cues, often perceiving them as typical signs of phishing attempts \cite{van2019cognitive}. Even though there is a slight increase usage of \textit{``urgent''} and \textit{``warning''} in Fig.~\ref{fig_iso_split_features}, our analysis reveals that PEEK anomalies strategically reduce the emphasis on urgency cues ($median_{nml} = 2.38$, $median_{anml} = 2.94$). As shown in Table \ref{tab:principle_frequent_words}, instead of overtly pressuring recipients, PEEK phishing utilizes terms like \textit{``before''}, \textit{ ``continue''}, and \textit{``time''}, which subtly extend time frame and mitigate the sense of immediate action. This more nuanced approach may increase the time readers spend engaging with the content, allowing other persuasive strategies to have a greater impact on decision-making \cite{xiao2019changing}. 

A further distinction lies in the use of \textit{Consistency}. 
The principle often involves creating commitments through agreements such as insurance policies or tax arrangements, emphasizing long-term stability and a forward-looking perspective \cite{van2019cognitive}. Normal phishing emails tend to be direct and assertive, using phrases such as \textit{``have to''} and \textit{``make sure''} to compel the recipient to take immediate action. In contrast, PEEK anomalies adopts a more tentative tone with words like \textit{``potential''}, \textit{``may''}, \textit{``will''}, and \textit{``ready to''}. The broader range of expression is associated with increased perceived authenticity, which raises the probability of being persuaded \cite{pennebaker2015development}.

Early research suggests that the combination of cognitive complexity with a calming emotional tone may increase the likelihood of victims moving from simple reading to deeper cognitive thinking
\cite{pennebaker2015development}. Our analysis reveals that traditional phishing emails rely on creating a sense of urgency and exploiting ambiguity, where PEEK anomalies tend to provide more detailed information while mitigating emotional tension. These contrasts underscore the necessity of updating detection mechanisms to effectively accommodate the evolving strategies employed in LLM-driven phishing attacks.


\begin{tcolorbox}[colback=gray!20,boxsep=1mm, left=0.125mm, right=0.125mm, top=0.125mm, bottom=0.125mm]
\textbf{Takeaway 4}: PEEK anomalies demonstrate significant divergence in expressing \textit{Scarcity} and \textit{Consistency}, adopting more emotionally neutral, authentic-sounding narratives. This makes LLM-driven phishing both more subtle and potentially more persuasive.
\end{tcolorbox}

\subsection{RQ3: Cyclic Process based on Extracted Phishing Patterns.}\label{chapter_RQ3} 
\begin{table*}[ht]
    \centering
    \caption{Model Performance under DeepWordBug Attack Using Different Training Datasets. Where \textit{EVA-Acc} and \textit{EVA-F1} are Accuracy and F1-score under evasion attacks; Attack Success Rate \textit{ASR}, indicating the percentage of data bypassing detectors; Comapred Synthetic Datasets:\protect\colorbox{yellow!30}{IWSPA\_2023}, 
    \protect\colorbox{green!20}{DeepAI}, \protect\colorbox{cyan!25}{PEEK phishing}.}
    \label{tab:evasion_attacks_three_datasets_4}
    \begin{threeparttable}
    \begin{tabular}{p{1cm}cccccp{2cm}cccccc}
        \hline
          Model & Acc & F1 & EVA-Acc & EVA-F1 & ASR\%& Model & Acc & F1 & EVA-Acc & EVA-F1 &ASR\% \\
        \hline
        \multicolumn{6}{l}{\textit{\textbf{$D_{1}$} (Existing Phishing + Benign)}} & \multicolumn{6}{l}{\textit{\textbf{$D_{2}$} (Existing Phishing + PEEK phishing)}} \\
        \hline
                &\cellcolor{yellow!30}0.63	&\cellcolor{yellow!30}0.71	&\cellcolor{yellow!30}0.17	&\cellcolor{yellow!30}0.09 &\cellcolor{yellow!30}74.38       &         
                &\cellcolor{yellow!30}0.64	&\cellcolor{yellow!30}0.70	&\cellcolor{yellow!30}0.33	&\cellcolor{yellow!30}0.27& \cellcolor{yellow!30}23.15       \\
        SQ  	&\cellcolor{green!20}0.55	&\cellcolor{green!20}0.62	&\cellcolor{green!20}0.64	&\cellcolor{green!20}\textbf{0.72} &\cellcolor{green!20}67.2	& T5-Encoder      &\cellcolor{green!20}0.55	&\cellcolor{green!20}0.60	&\cellcolor{green!20}0.21	&\cellcolor{green!20}0.18& \cellcolor{green!20}2.77\\
	        &\cellcolor{cyan!25}\textbf{0.90}	&\cellcolor{cyan!25}\textbf{0.91}	&\cellcolor{cyan!25}\textbf{0.70}	
            &\cellcolor{cyan!25}0.70 & \cellcolor{cyan!25}\textbf{18.76} &    &\cellcolor{cyan!25}\textbf{0.99}	&\cellcolor{cyan!25}\textbf{0.99}	&\cellcolor{cyan!25}\textbf{0.96}	&\cellcolor{cyan!25}\textbf{0.96} &\cellcolor{cyan!25}\textbf{2.13} \\
        \hline
                  &\cellcolor{yellow!30}0.59	&\cellcolor{yellow!30}0.74	&\cellcolor{yellow!30}0.49	&\cellcolor{yellow!30}\textbf{0.65}&	 \cellcolor{yellow!30} \textbf{17.85} &          &\cellcolor{yellow!30}0.52	&\cellcolor{yellow!30}0.68	&\cellcolor{yellow!30}0.19	&\cellcolor{yellow!30}0.32 &\cellcolor{yellow!30}10.67 \\
        ALBERT    &\cellcolor{green!20}0.58	&\cellcolor{green!20}0.77	&\cellcolor{green!20}0.21	&\cellcolor{green!20}0.40    &	\cellcolor{green!20} 87.6 & DistilBERT             &\cellcolor{green!20}0.52	 &\cellcolor{green!20}0.68	 &\cellcolor{green!20}0.52	 &\cellcolor{green!20}0.68 &\cellcolor{green!20}11.76 \\
                  &\cellcolor{cyan!25}\textbf{0.88}	&\cellcolor{cyan!25}\textbf{0.89}	&\cellcolor{cyan!25}\textbf{0.55}	&\cellcolor{cyan!25}0.57& \cellcolor{cyan!25}40.91 & &\cellcolor{cyan!25}\textbf{0.99}	&\cellcolor{cyan!25}\textbf{0.99}	&\cellcolor{cyan!25}\textbf{0.97}	&\cellcolor{cyan!25}\textbf{0.97} &\cellcolor{cyan!25}\textbf{2.42} \\
        \hline
                 & \cellcolor{yellow!30}0.83	&\cellcolor{yellow!30}0.83	&\cellcolor{yellow!30}0.53	&\cellcolor{yellow!30}0.41&	   \cellcolor{yellow!30} 26.51  &        &\cellcolor{yellow!30}0.47	&\cellcolor{yellow!30}0.64	&\cellcolor{yellow!30}0.42	&\cellcolor{yellow!30}0.59 &\cellcolor{yellow!30}2.77 \\
        RoBERTa  &\cellcolor{green!20}0.58 	&\cellcolor{green!20}0.74	&\cellcolor{green!20}0.57	&\cellcolor{green!20}0.56 &	 \cellcolor{green!20} 32.6    & ROBERTa-Syn    &\cellcolor{green!20}0.51 	&\cellcolor{green!20}0.67 	&\cellcolor{green!20}0.49 	 &\cellcolor{green!20}0.66 &\cellcolor{green!20}3.94 \\
                 &\cellcolor{cyan!25}\textbf{0.91}	&\cellcolor{cyan!25}\textbf{0.92}	&\cellcolor{cyan!25}\textbf{0.77}	&\cellcolor{cyan!25}\textbf{0.78}   &\cellcolor{cyan!25}\textbf{8.76}  & &\cellcolor{cyan!25}\textbf{0.99}	&\cellcolor{cyan!25}\textbf{0.99}	&\cellcolor{cyan!25}\textbf{0.87}	&\cellcolor{cyan!25}\textbf{0.99} &\cellcolor{cyan!25}\textbf{1} \\
        \hline
    \end{tabular}
    \end{threeparttable}
\end{table*}
\begin{table*}[th]
    \centering
    \caption{Model Performance under DeepWordBug Attack. \textbf{Left}: Performance before PEEK phishing Fine-Tuning. \textbf{Right}: Performance after PEEK phishing Fine-Tuning. \textit{EVA} represents Evasion Attacks, where \textit{EVA-Acc} records accuracy under such attacks, and \textit{ASR (Attack Successful Rate)} indicates the percentage of data that successfully bypassed the detectors.}
    \begin{tabular}{cc}
        \begin{minipage}{0.48\linewidth}
            \centering  
                \resizebox{\linewidth}{!}{
                \begin{tabular}{lccccc}
                \hline
                  \textbf{Model} & \textbf{Acc} & \textbf{F1} & \textbf{EVA-Acc} & \textbf{EVA-F1} & \textbf{ASR\%}\\
                    \hline
                \multicolumn{6}{l}{\textit{\textbf{$D_{1}$ (Existing Phishing + Benign)}}}\\
                \hline
                SQ & 0.32 & 0.42 & 0.21 & 0.15 & 36.81\\
                ALBERT & 0.54 & 0.57 & 0.18 & 0.13 & 67.15\\
                RoBERTa& 0.54 & 0.57 & 0.46 & 0.43 & 14.81\\
                \hline
                \multicolumn{6}{l}{\textit{\textbf{$D_{2}$ (Existing Phishing + PEEK phishing)}}}\\
                \hline
                T5-Encoder& 0.53 & 0.68 & 0.49 & 0.66 & 6.74 \\
                DistilBERT& 0.42 & 0.51 & 0.13 & 0.03 & 27.44\\
                RoBERTa-Syn& 0.44 & 0.35 & 0.16 & 0.09 & 64.86\\
                     \hline
                \end{tabular}
                }
                \label{tab:deepwordbug_attack_before}
        \end{minipage} &
        
        \begin{minipage}{0.48\linewidth}
            \centering  
                \resizebox{\linewidth}{!}{
                \begin{tabular}{lccccc}
                \hline
                  \textbf{Model} & \textbf{Acc} & \textbf{F1} & \textbf{EVA-Acc} & \textbf{EVA-F1} & \textbf{ASR\%}\\
                    \hline
                \multicolumn{6}{l}{\textit{\textbf{$D_{1}$ (Existing Phishing + Benign)}}}\\
                \hline
                SQ & 0.90 & 0.91 & 0.70 & 0.69 & 18.76\\
                ALBERT & 0.88 & 0.89 & 0.55 & 0.57 & 40.91\\
                RoBERTa& 0.91 & 0.92 & 0.77 & 0.78 & 8.76 \\
                \hline
                \multicolumn{6}{l}{\textit{\textbf{$D_{2}$ (Existing Phishing + PEEK phishing)}}}\\
                \hline
                T5-Encoder& 0.99 & 0.99 & 0.96 & 0.96 & 2.13\\
                DistilBERT& 0.99 & 0.99 & 0.97 & 0.97 & 2.42\\
                RoBERTa-Syn& 0.99 & 0.99 & 0.97 & 0.99 & 1\\
                     \hline
                \end{tabular}
                }
                \label{tab:deepwordbug_attack_after}
        \end{minipage} \\
    \end{tabular}
\end{table*}

Our prior findings from Section~\ref{chapter_RQ1} and Section~\ref{chapter_RQ2} demonstrated diversified attack landscapes with different persuasive strategies driven by LLMs. There is a question regarding the applicability to anti-phishing attacks (RQ3). We assess the effectiveness of generated PEEK phishing through the criteria and corresponding metrics we mentioned in Section~\ref{chapter_metrics_evaluation_metrics}. 

\subsubsection{\textbf{\textit{Effectiveness of PEEK phishing}}} To assess the effectiveness of PEEK phishing, we evaluate the performance of detectors on phishing detection and robustness against perturbation attacks,  before and after fine-tuning using PEEK phishing samples. We compare PEEK phishing against two existing LLM-generated phishing datasets (IWSPA\_2023 and DeepAI-phishing) using two detector categories: 1) ALBERT, RoBERTa, and SQ \cite{mehdi2023adversarial}, designed to classify human-written phishing and benign emails ($D_{1}$); and 2) DistilBERT, RoBERTa-Syn \cite{jamal2024improved}, and T5-Encoder \cite{bethany2024large}, designed to distinguish between human-written and LLM-generated phishing emails ($D_{2}$). Perturbation attacks were simulated using four textual attack frameworks: TextFooler \cite{wang2020infobert}, PWWS \cite{ren2019generating}, Pruthi \cite{pruthi2019combating}, and DeepWordBug \cite{gao2018black}. Performance metrics include  \textit{accuracy}, \textit{F1-score}, and \textit{attack success rate (ASR)}. The experiments are set as follows:
\begin{itemize}[leftmargin=*]
    \item Evaluate the performance of detectors on phishing detection and resilience against adversarial perturbation attacks;
    \item Fine-tune detectors using IWSPA\_2023 phishing, DeepAI-phishing, and PEEK phishing);
    \item Assess the performance of detectors after fine-tuning on phishing detection and defending against perturbation attacks.
\end{itemize}

Results indicate that detectors fine-tuned with PEEK phishing exhibit enhanced robustness compared to those trained on alternative datasets  (Table~\ref{tab:evasion_attacks_three_datasets_4} and Table~\ref {tab:evasion_attacks_three_datasets_1} to \ref{tab:evasion_attacks_three_datasets_3}). Following fine-tuning with PEEK phishing (Table \ref{tab:deepwordbug_attack_after}, right), $D_{1}$ models showed significant improvements; for instance, RoBERTa achieved an F1-score of 0.91 and reduced its ASR from 14.81\% to 8.76\%. 
Even the least effective model, ALBERT, lowered its ASR from 67.15\% to 40.91\% while achieving an F1-score of 0.88. For $D_{2}$, all detectors achieved near-perfect detection accuracy and F1-score (both are 0.99), along with strong resistance to adversarial attacks, with RoBERTa-Syn effectively mitigating 63\% of perturbation attacks. These findings underscore the efficacy of PEEK phishing in detector robustness and reducing vulnerability to adversarial perturbations. 
Further implementation details are provided in Appendix~\ref{appendix:evasion_attacks_other_results}.

\begin{tcolorbox}[colback=gray!20,boxsep=1mm, left=0.125mm, right=0.125mm, top=0.125mm, bottom=0.125mm]
\textbf{Takeaway 5}: PEEK phishing significantly enhances detector performance, increasing F1-scores by up to 88\% and reducing perturbation attack success rate by as much as 70\%.
\end{tcolorbox}

\subsubsection{\textbf{\textit{Transferability Across Phishing Topics}}} 
To evaluate PEEK's cross-topic phishing generation capabilities, we created phishing samples across diverse domains, including Document Sharing, Donation, Education, Job Hiring \& Business, Shipment \& Package \& Delivery, Tax \& Payroll, and Audio Message \cite{lund2023chatgpt}. Topic-specific keywords derived from ChatGPT (Appendix \ref{appendix:expand_education_topic_keywords}) balanced benign and phishing intents, generating 100 samples per topic. Analyzing PAS, 88.1\% of the cross-topic samples were classified as realistic phishing attempts, confirming PEEK's robust capability for high-quality zero-shot generation (examples provided in Appendix \ref{appendix:other_phishing_example}).
\begin{tcolorbox}[colback=gray!20,boxsep=1mm, left=0.125mm, right=0.125mm, top=0.125mm, bottom=0.125mm]
\textbf{Takeaway 6}: PEEK achieves high-quality phishing generation across varied phishing topics (PAS = 88.1\%).
\end{tcolorbox}

\begin{figure}[ht]
\centering
\includegraphics[width=\linewidth, clip, trim = 0 0 0cm 0.6cm]{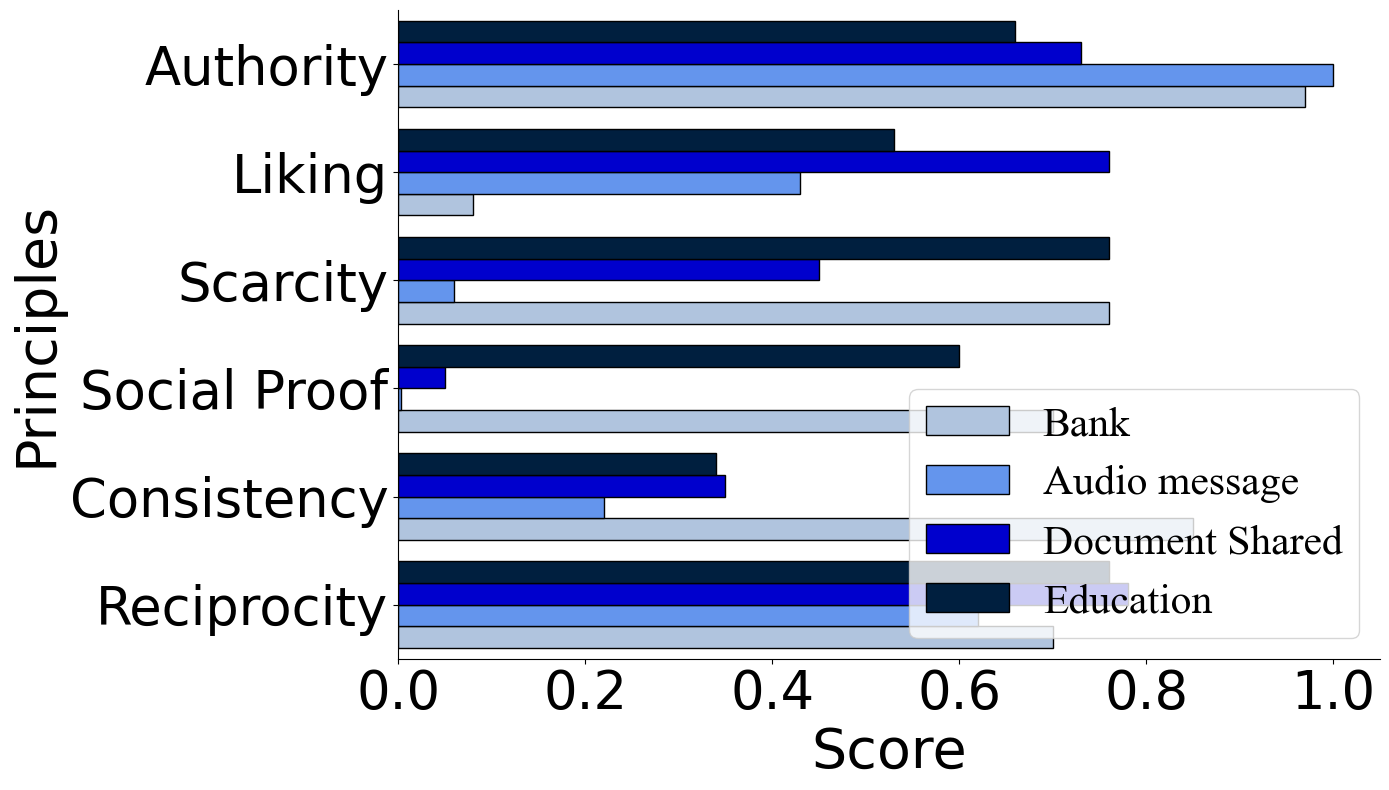}
\caption{Persuasion Principle Usage Across Phishing Themes. Higher scores indicate greater use of persuasive principles. Different domains exhibit distinct deceptive tactics.}
\label{fig_extra_principles_frequency}
\vspace{1mm}
\end{figure}

To investigate employed persuasion strategies across different topical phishing, we employed the DPS metric to quantify the principles presented in the samples. As illustrated in Fig.~\ref{fig_extra_principles_frequency} (the remaining in Fig.~\ref{fig_extra_principles_frequency_rest}), distinct email themes appear to exploit different human vulnerabilities. For example, phishing emails within educational-themed contexts commonly incorporate the principles of \textit{Reciprocity} and \textit{Scarcity}, using time-sensitive prompts to induce rapid responses. 
Conversely, audio phishing attacks predominantly leverage \textit{Authority}, enhancing perceived credibility and control. 

In summary, we found the principles of \textit{Reciprocity} and \textit{Authority} are commonly employed in PEEK-generated phishing emails, which reflects that LLMs primarily exploit desires and trust to challenge individuals' psychological defenses and social norms. The universality of human desires across different contexts enables phishers to launch widespread phishing attacks in terms of attack scale and across domains \cite{wang2021social}. Consequently, understanding such persuasive strategies is crucial for strengthening detection mechanisms and mitigating the risks posed by multi-theme phishing attacks. 
\begin{tcolorbox}[colback=gray!20,boxsep=1mm, left=0.125mm, right=0.125mm, top=0.125mm, bottom=0.125mm]
\textbf{Takeaway 7}: LLMs commonly leverage \textit{Authority} and \textit{Reciprocity} to facilitate generalized deception by manipulating universal desires. Identifying and countering these strategies is essential to mitigating widespread phishing attacks.
\end{tcolorbox}

\section{Discussion} \label{chapter_discussion}
We now discuss the limitations and suggest future work. 
\noindent \textbf{Preventing Misuse of PEEK.} In our work, PEEK was proposed to enhance the quality and diversity of phishing emails while understanding potential changes in deceptive patterns. However, the accessibility of the fine-tuned generator and prompts results in a risk of adversarial abuse. For instance, phishers might exploit the PEEK generator to facilitate cross-topic phishing campaigns. This highlights the need for robust countermeasures to mitigate complex and diverse phishing campaigns that exist independently from PEEK. We will publicly release PEEK fine-tuned detectors for PEEK-driven phishing defense to improve their robustness against strategies demonstrated by PEEK phishing. Future work will keep fine-tuning these detectors and might investigate complementary safeguards such as light watermarking \cite{zhang2020udh}, attribution strategies \cite{panum2020towards} to mitigate the direct risks of misuse of PEEK components. 

\noindent \textbf{Multi-Scope Phishing Features Analysis.} 
Our current evaluation is confined to plain-text emails. Yet modern phishing increasingly exploits visually oriented vectors, e.g., QR codes~\cite{geisler2024hooked} or image-based notifications \cite{wang2023image,li2024knowphish}.
Such multimodal attacks strain detectors that rely solely on textual cues. 
The growing diversification of phishing attacks continues to challenge detection systems, highlighting the need for more comprehensive and adaptive defense countermeasures. Future extension could explore multimodal phishing simulations, such as text-to-image, to leverage the text-based PEEK approach to other media types.

\noindent \textbf{Prompt-driven Attacks Analysis.} Rather than conventional data augmentation, we focus on using prompt-based strategies to enhance the phishing dataset's effectiveness. While prompts introduce greater flexibility, they can conflict with ethical generation constraints and might need to be designed to surpass such constraints. LLMs are exploited to update and refine malicious prompts automatically to generate target malicious content \cite{roy2024chatbots}. In this work, we examine one particular prompt formulation and its impact on phishing generation. Future research could systematically study how prompt constellations influence linguistic style, structure, and deception strategies will yield deeper insight into the boundaries of prompt-based defenses.

\noindent \textbf{Towards Future Phishing Analyzers.} 
Building on ChatGPT's capability in understanding and summarizing, ChatGPT has been used in cybersecurity threat analysis~\cite{charfeddine2024chatgpt}. In this work, we adopt ChatGPT based on OpenAI's \textit{gpt-3.5-turbo} model,
following the prompt from Koide et al.~\cite{koide2024chatspamdetector} to guide deeper phishing intent analysis within the PEEK framework. While more advanced models are released (e.g., GPT-4 and its variants), our 5-fold cross-validation yields high F1 and F2 scores, suggesting that \textit{gpt-3.5-turbo} provides sufficient contextual understanding and generalization for our phishing analysis scenario.
However, the analyzers will evolve alongside the threats they are designed to detect. Analyzers may harness techniques such as retrieving augmented generation (RAG) and multimodal detection, enabling dynamic adaptation to emerging threats. Future work may incorporate such advanced analyzers to resemble more realistic phishing samples.

\section{Related Work} \label{chapter_related_work}
\textbf{Generating High-Quality and Diverse Phishing Emails}. High-quality and diverse phishing datasets are central to current phishing defense countermeasures \cite{gong2023survey}. Effective datasets for defense training are both high-quality to accurately represent realistic phishing attacks, and diverse to cover a wide range of phishing components, such as topics \cite{das2019sok}. Previous studies have combined Feature Engineering (FE) and Deep Learning (DL) techniques to generate phishing emails~\cite{salloum2022systematic,alhogail2021applying,al2021generating,seymour2016weaponizing}. FE is used to extract phishing indicators, such as function words \textit{``click''} and \textit{``urgent''}; where DL models focus on learning the contextual patterns of these features to generate similar content. However, FE-based phishing generation is limited by insufficient feature identification, especially when representative features change, which reduces the utility of the provided phishing content \cite{das2019sok}. Differently, Gholampour et al. generated phishing emails by applying data augmentation techniques using text augmenters such as Textfooler \cite{jin2020bert}, which introduces slight text variations by adding, replacing, or removing words while preserving meaning \cite{mehdi2023adversarial}. This approach improved the availability of phishing emails from minority classes and enhanced content diversity through the perspective of perturbative words. 

Recently, large language models (LLMs) have been implemented in phishing email generation, benefiting from their large-scale pre-training corpus and advanced generative capabilities \cite{koide2024chatspamdetector,chen2024survey}. Previous studies commonly combine prompt engineering with LLMs for phishing generation, using prompt engineering to craft instructions that guide the generation of target content. Recent research has explored both zero-shot prompting \cite{mehdi2023adversarial} and few-shot prompting \cite{bethany2024large,roy2024chatbots,eze2024analysis} for phishing generation. Bethany et al. \cite{bethany2024large} investigate the quality of LLM-generated phishing emails in a real-world setting involving 9,000 employees during 11 months. The study crafts scenario-specific prompts to instruct LLMs in generating phishing emails. They evaluate LLM-generated phishing quality by measuring the percentage of participants who provide private information, and compare this with the results from human-written phishing emails. A higher percentage indicates higher-quality phishing datasets, as it reflects the presence of novel spear-phishing tactics. Roy et al. \cite{roy2024chatbots} proposed a closed framework that automatically generates phishing scams according to LLMs' self-generated prompts. The automated process includes collecting data, generating content, providing feedback, and regenerating improved phishing emails. The last three steps aim to iteratively improve phishing email quality by having the LLM rewrite content using prompts that it refines through self-feedback. These studies primarily focus on generating phishing emails leveraging designed prompts tailored to specific attack scenarios, mainly assessing the phishing generation capabilities of LLMs and exploring the associated security risks.\textit{ Unlike these studies, our work employs LLMs to generate a wide range of phishing emails using chat-based prompts to increase content diversity, while improving email quality through an adversarial training approach. This work aims to provide high-quality and diverse synthetic phishing content to enhance detector performance against adaptive attacks.}

\noindent \textbf{Persuasion Principles in Phishing Attacks}.
Social Engineering in Cybersecurity (SEiCS) has been defined as an attack method where attackers exploit human psychological vulnerabilities through social interactions to compromise cybersecurity \cite{hatfield2018social}. The social interaction enables phishers to manipulate victims, for example, by sending phishing emails~\cite{wang2020defining}. Human vulnerabilities involve cognitive bias, behavioral habits, emotional response, human nature, and personality traits, encompassing six aspects of individual disposition \cite{wang2021social}. These vulnerabilities are reflected in Cialdini's persuasion principles~\cite{cialdini2001science}, namely ``\textit{reciprocity}'', ``\textit{consistency}'', ``\textit{social proof}'', ``\textit{authority}'', ``\textit{liking}'', and ``\textit{scarcity}'', which are widely used to offer insights into the design of effective defense strategies~\cite{Kashapov22email}.

The study conducted by Lawson et al. \cite{lawson2020email} found that people who are extroverted are more susceptible to phishing. This is because they are more likely to build trust and respond to positive emotional cues from phishers, especially tactics crafted by principles of \textit{``Liking''} and \textit{``Scarcity''}, which together amplify the perceived urgency and value of limited opportunities. Accordingly, the persuasive principles can also be integrated into machine learning to predict the success or failure of phishing attacks. Van Der Heijden and Allodi \cite{van2019cognitive} quantified the principles in phishing emails and found that more cognitive weaknesses attackers exploited, the higher the probability of users being deceived. One common cognitive weakness is the principle of \textit{``scarcity''}, which manipulates users into responding impulsively and falling into the trap. However, recent studies show that participants are more likely to label an email as phishing because they are familiar with common expressions of \textit{Authority} and \textit{Scarcity}, such as “must call us immediately” \cite{li2020detection}. Earlier research has primarily concentrated on the psychological dynamics of phishing in human-to-human (H2H) interactions; however, with the emergence of LLMs, the interaction pattern has shifted to machine-to-human (M2H) \cite{wang2020defining, Quan24email}. This transition has accelerated the evolution of phishing attacks and introduced new, distinct characteristics \cite{schmitt2023digital}. As a result, features derived from traditional H2H interaction models may no longer fully capture the characteristics of phishing content generated by LLMs~\cite{salahdine2019social}, which requires enhancing the transparency of the M2H module \cite{nguyenai2tale}. \textit{Unlike these studies, we investigate the M2H interactions featured in phishing campaigns by leveraging LLMs to mimic realistic phishing attacks. Based on the generated content, we further explore how deceptive strategies differ from those used in H2H interactions.}

\section{Conclusion}\label{chapter_conclusion}
Phishing continues to be a widespread cyber threat, with attackers using deceptive emails to steal sensitive information. Although AI, especially deep learning and large language models, plays a critical role in defending against phishing, the lack of diverse, up-to-date data due to privacy concerns limits its effectiveness, making it difficult for models to detect new and sophisticated phishing tactics. We propose the Phishing Evolution Framework (PEEK), which is an iterative framework that leverages LLMs to support diverse and quality phishing generation and track the evolving tactics of LLM-generated phishing samples. The framework can produce high-quality, diverse phishing samples, enhancing the robustness of current detectors against adversarial attacks. Additionally, we incorporate LLM-based analysis and linguistic analysis to gain a deeper understanding of the differences between LLM-generated phishing and existing phishing datasets.

\bibliographystyle{IEEEtran}
\bibliography{sample-base}

\begin{appendices}
\section{Related Dataset Information}
\subsection{Existing Phishing Dataset Information}\label{appendix:dataset_info}
We present information on our collected datasets in Table~\ref{datasets_information}, accompanied by brief descriptions. Specific acquisition methods can be found through relevant references. In total, we collected six phishing datasets and one benign email dataset. The emails span a broad timestamp, covering publicly available datasets from 1998 to 2021. It is noteworthy that an email might appear multiple times, as different sources employ varying storage practices. Data pre-processing was essential, including deduplication, adjusting email lengths, removing highly similar emails, and standardizing storage formats. Dataset information after processing can be referred to as Table \ref{tab:dataset_distribution}.

\begin{table}[ht]
\caption{Collected Existing Dataset Information}
\resizebox{\columnwidth}{!}{
  \begin{tabular}{p{1.5cm}p{6.5cm}}
    \hline
    Datasets&Description\\
\hline
IWSPA-AP (2018) &
The dataset was collected as part of a shared task to try to address phishing scam emails. The dataset has a combination of benign and malicious email bodies. \\
\hline
Nazario Phishing Corpus &
Dataset of phishing emails received by one user. The emails are surrounded by HTML code and would need to be stripped to get the useful text data.\\
\hline
Miller Smiles &
The dataset contains over 2.5 million ``scam'' emails dated from 2003 to 2021. The dataset contains the bodies of phishing email scams.\\
\hline
Phish Bowl Cornell University&
The dataset contains phishing emails that have been spotted and reported by students and staff at Cornell University. The dataset has been sanitized to some extent. It contains over 2000 entries.\\
\hline
CLAIR collection of ``Nigerian'' fraud emails&
This Kaggle dataset contains more than 2,500 fraudulent emails attempting Nigerian Letters or ``419'' Fraud. All the emails are in one text file and contain a large amount of header data. These emails were collected between 1998 to 2007.\\
\hline
Cambridge&
This dataset contains a large number of email headers, involving information such as sending and receiving addresses, email subjects, etc. It also includes the body content of phishing emails.\\
\hline
Enron Email Dataset [Benign]&
Large email dataset of 0.5M emails that were released as part of an investigation into Enron. It consists of emails from 150 users, mostly senior management of Enron.\\
  \hline
\end{tabular}
\label{datasets_information}
}
\end{table}

\begin{table}[ht]
\small
    \centering
    \caption{Tabular Representation of Phishing Datasets Quantities Before and After Data Wrangling. Description of each dataset can be found in Appendix \ref{appendix:dataset_info}.}
    \resizebox{\columnwidth}{!}{
    \begin{tabular}{ccc}
        \hline
        \textbf{Dataset}&\textbf{Before Processing}&\textbf{After Processing}\\
        \hline
        IWSPA-AP \cite{IWSPA}&  628&123 \\
        Nazario \cite{Nazario}& 946&292\\
        Miller Smiles \cite{Miller_Smiles} &23366 &5021\\
        Phishing Bowl \cite{phish_bowl}&1757&435\\
        Nigerian Fraud \cite{Nigerian_Fraud}& 3792&857\\
        Cambridge Phishing& 41446&7071\\
     \hline
    \end{tabular}
    }
    \label{tab:dataset_distribution}
\end{table}

\subsection{Generated Examples for Phishing Analysis}\label{appendix:other_phishing_example}
\begin{figure}[H]
  \centering
  \subfloat[Topic Across sample from PEEK labeled as phishing.]{%
    \includegraphics[width=\linewidth]{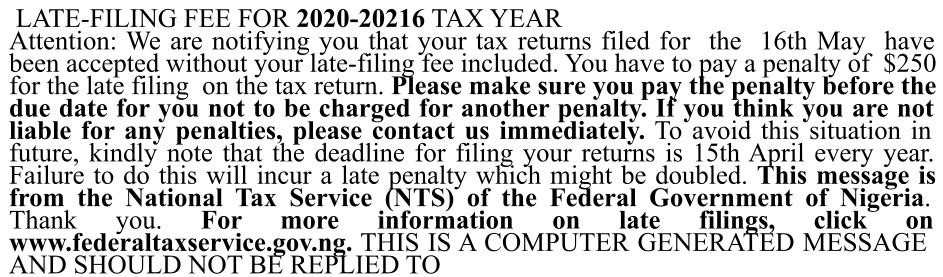}%
    \label{fig_other_phishing}%
  }\\
  \subfloat[Topic Across sample from PEEK labeled as non-phishing.]{%
    \includegraphics[width=\linewidth,clip,trim=0cm 0cm 0cm 0cm]{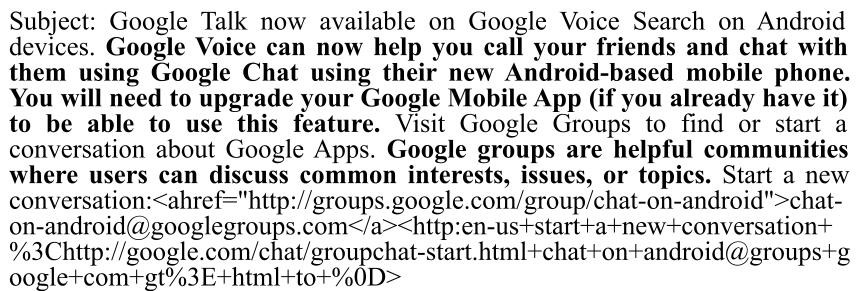}%
    \label{fig_other_non_phishing}%
  }
  \caption{Examples of PEEK-generated Email Sample}
  \label{fig_other_topic_samples}
\end{figure}

\section{Object Function and Model Configuration}
\subsection{Object Function}\label{appendix:object_function}
In this work, we denote the dataset as $D = \{(\bm{x}_1,y_1),\cdots,(\bm{x}_{n}, y_{n})\}$, where $\bm{x}_i$ represents the email content, $y_i$ is the label of email $\bm{x}_i$ where $y_{i}\in\{0,1\}$ ($0$ is benign and $1$ is phishing), and $n$ is the number of data samples. We integrated labeled, processed phishing data into a custom chat prompt to train the generator, which was designed to produce highly realistic phishing samples.

In the first phase of our PEEK, we train the generator (LLM-G, or $G$ for short) and discriminator (D) through adversarial training (via the GAN principle). This competitive process mutually improves their performance, ultimately enhancing the generator's generative capability \cite{goodfellow2014generative}. 
To be specific, the generator $G(\cdot)$ learns to capture the data distribution, while the discriminator $D(\cdot)$ focuses on estimating the probability that a sample came from the training data in place of $G$. The generator $G$ learns a generator distribution $p_g$ used for generating realistic phishing samples. It builds a mapping function from a prior distribution $p_z$, conditioned on a template prompt, to the data space $G(z)$. In the meantime, the discriminator $D$ outputs a scalar representing the probability that $\bm{x}$ came from training data (i.e., $p_{data}$) rather than $p_g$ \cite{goodfellow2014generative}.

As mentioned in Section \ref{chapter_pen_A}, a template prompt has been designed to optimize $G(\cdot)$ net performance. As the training data consists of benign and phishing content, we leverage label $y$ as assistant information and embed it into \textit{Instructions} of prompts, clarifying and guiding generations. To be more specific, we use the label $y$ (of a specific data $\bm{x}$) by embedding its information (i.e., phishing or benign) into a corresponding prompt. We denote the prompt as $I_{g}$, which serves as the input to the $G(\cdot)$. With the label class information, $G(\cdot)$ net is guided to generate more targeted samples. In our proposed framework, during this training process, the generator $G$ and the discriminator $D$ are trained simultaneously via the following objective function:

\begin{equation}
\min_{G}\max_{D} \ \ E_{\bm{x}\sim p_{data}}[logD(\bm{x})]+E_{\bm{z}\sim p_{z}}[log(1-D(G(I_{g})))]\label{eq:v1}
\end{equation}

In the objective function shown in Eq.(\ref{eq:v1}), the goal of $G(\cdot)$ net is to generate text that resembles real data and corresponds to label $y$ (i.e. if label $y$ is phishing, the generator $G(\cdot)$ is trained to generate real phishing data). In particular, the generator minimizes its loss by updating its parameters to generate more realistic data, and the discriminator minimizes its loss by updating its parameters to become better at distinguishing real and generated data. In the end, after the training phase, we get the generator $G$ with applicability on label-instruction text generation missions. The above networks, $G(\cdot)$ and $D(\cdot)$ are often non-linear mapping functions \cite{mirza2014conditional}, and we employed Llama 3.1 8B \cite{Llama-3.1} in $G$ net.
Simultaneously harness Bi-LISM Attention as $D$ net. During the training phase, the generator and discriminator networks will be updated simultaneously. In the end, when the training process stabilizes, the generator is capable of generating human-like phishing data similar to that from real phishing datasets.

\subsection{Model Configuration}\label{appendix:model_config}
In each phase of PEEK, different LLMs are employed for various tasks. Model configurations are described in detail:
\noindent \textbf{Models for LLM-Phishing Generation.} An ablation study among SOTA LLMs generators (see Appendix \ref{appendix:chapter_prompts_ablation_study}) led to the selection of Llama 3.1 for the phishing generation task. To optimize memory usage without sacrificing performance, we applied 4-bit quantization. LoRA tuning was performed with \textit{r = 16}, and \textit{alpha = 16}, updating only 8\% parameters of the model. Fine-tuning was conducted with a learning rate of \textit{3e-4}. The discriminative network is a bi-directional LSTM, configured with an \textit{input dimension} of 1024 and 256 neurons in \textit{hidden layer}. The final generator model was saved after 5 \textit{epochs} of training with GAN. All LLMs' training was carried out on an NVIDIA H100 GPU, which features 93.25 GB of memory (VRAM).

\noindent \textbf{Models for Evasion Attacks.} 
We employed two categories of deep learning models to simulate real-world phishing detection and evasion attacks: 1) Phishing Detection Models: ALBERT, RoBERTa, SQ \cite{mehdi2023adversarial} were chosen to classify legitimate and phishing emails; 2) Human vs. LLM-written Phishing Detection Models: DistilBERT, RoBERTa-LLMs \cite{jamal2024improved} and T5-Encoder \cite{bethany2024large} were used to differentiate between human-crafted and LLM-generated phishing emails. For the first stage of evasion attacks, we obtained initial model parameters from the respective original author. During the second stage of fine-tuning with filtered PEEK-generated datasets, we performed a grid search to identify the optimal learning rate from the set \textit{$\{3e-4, 1e-5, 2e-5, 3e-5, 5e-5\}$}. Due to space constraints, the final hyperparameters are available in the Appendix \ref{appendix:EVA_model_parameters}. All fine-tuning phases and imitation attacks were executed using an NVIDIA H100 GPU.

\subsection{ChatGPT Analyzer Cross-Validation}\label{appendix:analyzer_CV}
In recent years, LLMs (e.g., GPTs, Claude, Llama) have been commonly used in cybersecurity across diverse research areas, which plays a critical role in the process of cybersecurity framework \cite{charfeddine2024chatgpt}. Their functions contribute to \textit{Identify} risks, \textit{Protect} service delivery, \textit{Detect} cyber occurrences, \textit{Respond} to detection results, and \textit{Recover} disrupted results. Specifically to the role of defense, 
studies that have explored their application in security note taking~\cite{charfeddine2024chatgpt}, hate speech detection~\cite{das2023evaluating,huang2023chatgpt}, phishing intention analysis~\cite{koide2024chatspamdetector}, and so forth. The extensive training corpus and powerful learning structures enable the pre-trained or fine-tuned models to perform well in the aforementioned malicious analysis tasks. For instance, Koide et.al studied the performance of ChatGPT in phishing email analysis with well-crafted prompts, which achieves an accuracy of 99.70\%. The crafted prompts instruct analysis from the respective features of the email header, main body, and insert attachments \& URLs with highlighted content supporting results. The prompts result in outputs of \textit{is\_phishing}, a boolean value indicating the type of given emails; \textit{phishing\_score} quantifies the suspicion of phishing; and \textit{rationals} are to support determination.

To evaluate how closely PEEK-generated phishing to human-written phishing, we employed a representative LLM, ChatGPT, as an analyzer to validate the authenticity of PEEK-generated phishing samples inspired by the proficient contextual interpretation of LLMs. To evaluate the effectiveness of ChatGPT in phishing email analysis, we follow the prompt in Koide's guiding analysis. Totally 1,000 topic-representative phishing emails were sampled from the clustered benign and phishing datasets. We split this dataset into 5 subsets with balanced types of emails for 5-fold cross-validation (CV), where CV is commonly used to assess the generalization and robustness of predictive models~\cite{browne2000cross}. Detection performance on each folder is evaluated using confusion metrics (precision, recall, accuracy, and F1-score), and the F2-score, which emphasizes models' ability to capture positive cases while tolerating errors of false positives. Our results demonstrate that ChatGPT is reasonably applied in phishing email analysis with an average F1-score of 0.95 and F2-score of 0.97 (Table~\ref{tab:ChatGPT_CV_results}). 
\begin{table}[H]
    \centering
    \caption{Analyzer Cross-Validation 5-Fold Results}
    \resizebox{\columnwidth}{!}{
    \begin{tabular}{cccccc}
        \hline
        Folder & Precision & Recall & Accuracy & F1-score & F2-score \\
        \hline 
         1  &  0.92  &  0.97  &  0.94  &  0.91  &  0.97  \\
         2  &  0.93  &  0.97  &  0.95  &  0.93  &  0.97  \\
         3  &  0.97  &  0.96  &  0.97  &  0.97  &  0.96  \\
         4  &  0.93  &  0.97  &  0.95  &  0.92  &  0.97  \\
         5  &  0.99  &  0.96  &  0.97  &  0.99  &  0.96  \\
        \hline     
    \end{tabular}
    } 
    \label{tab:ChatGPT_CV_results}
\end{table}

\section{Ablation Study}\label{appendix:chapter_prompts_ablation_study}
\noindent \subsection{Generator Selection.} To evaluate the performance of different LLMs on text generation tasks, we compared three versions of Llama 3.1 8B: 1) Pre-trained Llama 3.1 8B (baseline), 2) Fine-tuned Llama 3.1 8B, and 3) PEEK-adversarially trained Llama 3.1. We also included a fine-tuned GPT-2 \cite{radford2019language} for comparison. All generators were fine-tuned using the same 80\% training data, learning features based on the chat template (Fig.~\ref{fig_llama_prompt_example}). The phishing email generation tasks involved both existing and extended topics, with PEEK phishing used as input to verify the authenticity of the emails.

\begin{table}[h]
    \centering
    \caption{Performance of Diverse Generators on Different Phishing Generation Cases. Training prompts are used on these models as shown in Fig.~\ref{fig_llama_prompt_example}. More generated results are classified as phishing, the better the model performs, and the higher the utilization of PEEK.}
    \resizebox{\columnwidth}{!}{
    \begin{tabular}{lcccccc}
    \hline
         & \multicolumn{3}{c}{Existing Topics} & \multicolumn{3}{c}{Extra Topics}\\
         \hline
         & FT-NLU\% & FT-Llama3.1\% & Analyzer\% & FT-NLU\% & FT-Llama3.1\% & Analyzer\%\\
         \hline
     PT-Model-8B & 20.6 & 22.3 & 21.4 & 33.6 & 42.7& 12.6 \\
     FT-Model-8B & 77.3 & 81.4 & 80.6 & 49.7 & 60.7& 58.1 \\
     FT-GPT-2 & 78.1 & 94.2 & 76.2 & 25.4 & 33.6& 32.3 \\
      \textbf{PEEK} & \textbf{88.2} & \textbf{99.6} & \textbf{84.8} & \textbf{57.4} & \textbf{67.5}& \textbf{88.1} \\
     \hline
    \end{tabular}
    } 
    \label{tab:generators_selection_filters_results}
\end{table}

Table \ref{tab:generators_selection_filters_results} shows the number of LLM-generated phishing samples considered realistic across different models. We found that fine-tuned LLMs performed much better than the pre-trained models. Except for the ChatGPT analyzer, we use collected phishing emails to fine-tune another two models (IBM NLU \cite{IBM-Cloud} and Llama 3.1 classifiers) for comparison. Among the fine-tuned models, GPT-2 outperformed Llama 3.1 8B in phishing generation tasks, with a slight 0.8\% improvement in IBM NLU (Table \ref{tab:generators_selection_filters_results}). This is due to GPT-2’s focus on generating text closely resembling the training data, though it struggles with phishing diversity (25.4\% in IBM NLU), highlighting challenges in transferring to varied data. When focusing on the Llama 3.1 models, results show that adversarial training improves performance, with the highest filter success rates of 88.2\% and 57.4\% in IBM NLU across two phishing domains. This confirms the importance of the GAN principle for improving LLM-generated phishing quality and generalization across different data types. Additionally, ChatGPT was used to objectively assess generator performance, where the PEEK model showed more stable results.

\noindent \subsection{Prompts Design.} We validate the effectiveness of our designed prompts using the analyzer described in Section \ref{chapter_pen_B} and two other fine-tuned classification models. Fig.~\ref{fig_llama_prompt_example} shows the conceptual design of the prompt used to guide phishing generation. To identify the best chat template instructions for training, we conducted three experiments to find the most effective prompts.
\begin{itemize}
    \item \textbf{Exp1.} Llama 3.1 8B Fine-Tune + Instructions + [Existing Data];
    \item \textbf{Exp2.} Llama 3.1 8B Fine-Tune + Instructions + C-Keywords + [Existing Data]
    \item \textbf{Exp3 (Ours).} Llama 3.1 8B Fine-Tune + Instructions + T-Keywords + [Existing Data];
\end{itemize}

\noindent \textit{``Existing Dataset''} that is surrounded by a bracket means it will not attend the inference stage after training. \textit{``T''} and \textit{``C''} before ``\textit{keywords}'' represent \textit{Topic} and \textit{Content} respectively. 
Each experiment follows the same process: fine-tuning, generation, and filtering. The number of generated LLM phishing samples is fixed using the same test dataset. We use the proportion from the filters to select the best prompt, helping the generator learn the features. The results in Table \ref{tab:prompts_ablation_filters_results} show the proportion of LLM-generated ``real'' phishing. We found that Exp.3 produces the highest percentage of realistic phishing, mainly due to the inclusion of dominant keywords. Compared to Exp.2, providing keywords improves output effectiveness (over 12\% of IBM NLU detections are authenticated as real phishing) and limits random outputs like dialogues between \textit{``Assistant''} and \textit{``Users''}. Additionally, a small decreased proportion of feedback from the filters when evaluating the output from Exp2. indicates that keywords provided for the inference stage do not have to be confined to the email topics. A bag of words (BOW) that is related to textual information is also good for LLM-phishing generation. For the time being, phishing samples that go through all of the stages of PEEK are from Exp3, which supports desirable results for both quality and efficiency. 
\begin{table}[ht]
    \centering
    \caption{Prompts Ablation Study Results of Selected Generator (Providing PEEK phishing Only).}
    \footnotesize
    \begin{tabular}{cccc}
    \hline
         & IBM NLU\% & Llama 3.1\%& Analyzer\\
         \hline
    \textbf{Exp1.} & 76.2 & 84.3&70.5 \\
    \textbf{Exp2.} & 82.7 & 94.4&82.7 \\
    \textbf{Exp3.} & \textbf{88.2} & \textbf{99.6}& \textbf{84.8}\\
    \hline
    \end{tabular}
    \label{tab:prompts_ablation_filters_results}
\end{table} 

\section{Evasion Attacks Results}
\subsection{Datasets Trained Detector Performance Under Evasion Attacks Comparison}\label{appendix:evasion_attacks_other_datasets}
\begin{table*}[ht]
    \centering
    \begin{threeparttable}
    \caption{Model Performance under TextFooler Attack Using Different Training Datasets.}
    \label{tab:evasion_attacks_three_datasets_1}
    \begin{tabular}{p{1cm}ccccc p{2cm}ccccc}
        \hline
        Model & Acc & F1 & EVA-Acc\tnote{1} & EVA-F1\tnote{2} & ASR\%\tnote{3} & Model & Acc & F1 & EVA-Acc & EVA-F1 & ASR\% \\
        \hline
        \multicolumn{6}{l}{\textit{\textbf{$D_{1}$} (Existing Phishing + Benign)}} & \multicolumn{6}{l}{\textit{\textbf{$D_{2}$} (Existing Phishing + PEEK phishing)}} \\
        \hline
              &\cellcolor{yellow!30}0.65 &\cellcolor{yellow!30}0.71 &\cellcolor{yellow!30}0.03 &\cellcolor{yellow!30}0.02 &\cellcolor{yellow!30}95.90 &
              &\cellcolor{yellow!30}0.64 &\cellcolor{yellow!30}0.70 &\cellcolor{yellow!30}0.46 &\cellcolor{yellow!30}0.32 &\cellcolor{yellow!30}37.82 \\
        SQ    &\cellcolor{green!20}0.55 &\cellcolor{green!20}0.62 &\cellcolor{green!20}0.09 &\cellcolor{green!20}0.17 &\cellcolor{green!20}89.52 &
               T5-Encoder &\cellcolor{green!20}0.55 &\cellcolor{green!20}0.60 &\cellcolor{green!20}0.54 &\cellcolor{green!20}0.58 &\cellcolor{green!20}23.15  \\
              &\cellcolor{cyan!25}\textbf{0.88} &\cellcolor{cyan!25}\textbf{0.89} &\cellcolor{cyan!25}\textbf{0.88} &\cellcolor{cyan!25}\textbf{0.77} &\cellcolor{cyan!25}\textbf{19.63} &
              &\cellcolor{cyan!25}\textbf{0.99} &\cellcolor{cyan!25}\textbf{0.99} &\cellcolor{cyan!25}\textbf{0.89} &\cellcolor{cyan!25}\textbf{0.90} &\cellcolor{cyan!25}\textbf{3.11} \\
        \hline
              &\cellcolor{yellow!30}0.59 &\cellcolor{yellow!30}0.74 &\cellcolor{yellow!30}0.21 &\cellcolor{yellow!30}0.34 &\cellcolor{yellow!30}64.65  &
              &\cellcolor{yellow!30}0.52 &\cellcolor{yellow!30}0.68 &\cellcolor{yellow!30}0.32 &\cellcolor{yellow!30}0.26 &\cellcolor{yellow!30}46.74  \\
        ALBERT &\cellcolor{green!20}0.58 &\cellcolor{green!20}0.74 &\cellcolor{green!20}0.54 &\cellcolor{green!20}\textbf{0.70} &\cellcolor{green!20}78.91 &
               DistilBERT &\cellcolor{green!20}0.52 &\cellcolor{green!20}0.68 &\cellcolor{green!20}0.52 &\cellcolor{green!20}0.68 &\cellcolor{green!20}37.82  \\
              &\cellcolor{cyan!25}\textbf{0.88} &\cellcolor{cyan!25}\textbf{0.89} &\cellcolor{cyan!25}\textbf{0.55} &\cellcolor{cyan!25}0.56 &\cellcolor{cyan!25}\textbf{49.55}  &
              &\cellcolor{cyan!25}\textbf{0.99} &\cellcolor{cyan!25}\textbf{0.99} &\cellcolor{cyan!25}\textbf{0.87} &\cellcolor{cyan!25}\textbf{0.88} &\cellcolor{cyan!25}\textbf{12.68}  \\
        \hline
              &\cellcolor{yellow!30}0.86 &\cellcolor{yellow!30}0.87 &\cellcolor{yellow!30}0.26 &\cellcolor{yellow!30}0.01 &\cellcolor{yellow!30}69.48  &
              &\cellcolor{yellow!30}0.52 &\cellcolor{yellow!30}0.68 &\cellcolor{yellow!30}0.45 &\cellcolor{yellow!30}0.62 &\cellcolor{yellow!30}10.67  \\
        RoBERTa &\cellcolor{green!20}0.46 &\cellcolor{green!20}0.68 &\cellcolor{green!20}0.58 &\cellcolor{green!20}0.72 &\cellcolor{green!20}58.49 &
                ROBERTa-Syn &\cellcolor{green!20}0.51 &\cellcolor{green!20}0.67 &\cellcolor{green!20}0.46 &\cellcolor{green!20}0.63 &\cellcolor{green!20}9.45   \\
              &\cellcolor{cyan!25}\textbf{0.90} &\cellcolor{cyan!25}\textbf{0.91} &\cellcolor{cyan!25}\textbf{0.72} &\cellcolor{cyan!25}\textbf{0.74} &\cellcolor{cyan!25}\textbf{19.60} &
              &\cellcolor{cyan!25}\textbf{0.99} &\cellcolor{cyan!25}\textbf{0.99} &\cellcolor{cyan!25}\textbf{0.94} &\cellcolor{cyan!25}\textbf{0.94} &\cellcolor{cyan!25}\textbf{5.42}\\
        \hline
    \end{tabular}
    \begin{tablenotes}
        \item[1] \textit{EVA-Acc}: Accuracy under evasion attacks.
        \item[2] \textit{EVA-F1}: F1-Score under evasion attacks.
        \item[3] \textit{ASR}: Attack Success Rate, indicating the percentage of data bypassing detectors.
        \item[4] \textit{\colorbox{yellow!30}{IWSPA\_2023}, \colorbox{green!20}{DeepAI}, \colorbox{cyan!25}{PEEK phishing}}
    \end{tablenotes}
    \end{threeparttable}
\end{table*}

\begin{table*}[ht]
    \centering
    \begin{threeparttable}
    \caption{Model Performance under PWWS Attack Using Different Training Datasets.}
    \label{tab:evasion_attacks_three_datasets_2}
    \begin{tabular}{p{1cm}ccccc p{2cm}ccccc}
        \hline
        Model & Acc & F1 & EVA-Acc & EVA-F1 & ASR\% & Model & Acc & F1 & EVA-Acc & EVA-F1 & ASR\% \\
        \hline
        \multicolumn{6}{l}{\textit{\textbf{$D_{1}$} (Existing Phishing + Benign)}} & \multicolumn{6}{l}{\textit{\textbf{$D_{2}$} (Existing Phishing + PEEK phishing)}} \\
        \hline
              &\cellcolor{yellow!30}0.63 &\cellcolor{yellow!30}0.71 &\cellcolor{yellow!30}0.09 &\cellcolor{yellow!30}0.07 &\cellcolor{yellow!30}87.01 &
              &\cellcolor{yellow!30}0.64 &\cellcolor{yellow!30}0.70 &\cellcolor{yellow!30}0.65 &\cellcolor{yellow!30}0.68 &\cellcolor{yellow!30}15.34 \\
        SQ    &\cellcolor{green!20}0.55 &\cellcolor{green!20}0.62 &\cellcolor{green!20}0.28 &\cellcolor{green!20}0.31 &\cellcolor{green!20}56.28 &
              T5-Encoder &\cellcolor{green!20}0.55 &\cellcolor{green!20}0.60 &\cellcolor{green!20}0.54 &\cellcolor{green!20}0.57 &\cellcolor{green!20}\textbf{8.53} \\
              &\cellcolor{cyan!25}\textbf{0.88} &\cellcolor{cyan!25}\textbf{0.89} &\cellcolor{cyan!25}\textbf{0.55} &\cellcolor{cyan!25}\textbf{0.52} &\cellcolor{cyan!25}\textbf{38.98} &
              &\cellcolor{cyan!25}\textbf{0.99} &\cellcolor{cyan!25}\textbf{0.99} &\cellcolor{cyan!25}\textbf{0.88} &\cellcolor{cyan!25}\textbf{0.88} &\cellcolor{cyan!25}11.45 \\
        \hline
              &\cellcolor{yellow!30}0.59 &\cellcolor{yellow!30}0.74 &\cellcolor{yellow!30}\textbf{0.35} &\cellcolor{yellow!30}\textbf{0.52} &\cellcolor{yellow!30}\textbf{40.40} &
              &\cellcolor{yellow!30}0.52 &\cellcolor{yellow!30}0.68 &\cellcolor{yellow!30}0.51 &\cellcolor{yellow!30}0.68 &\cellcolor{yellow!30}23.09 \\
        ALBERT &\cellcolor{green!20}0.58 &\cellcolor{green!20}0.74 &\cellcolor{green!20}0.12 &\cellcolor{green!20}0.11 &\cellcolor{green!20}76.94 &
               DistilBERT &\cellcolor{green!20}0.52 &\cellcolor{green!20}0.68 &\cellcolor{green!20}0.51 &\cellcolor{green!20}0.62 &\cellcolor{green!20}15.89 \\
              &\cellcolor{cyan!25}\textbf{0.88} &\cellcolor{cyan!25}\textbf{0.89} &\cellcolor{cyan!25}0.31 &\cellcolor{cyan!25}0.41 &\cellcolor{cyan!25}65.16 &
              &\cellcolor{cyan!25}\textbf{0.99} &\cellcolor{cyan!25}\textbf{0.99} &\cellcolor{cyan!25}\textbf{0.87} &\cellcolor{cyan!25}\textbf{0.88} &\cellcolor{cyan!25}\textbf{12.27} \\
        \hline
              &\cellcolor{yellow!30}0.83 &\cellcolor{yellow!30}0.83 &\cellcolor{yellow!30}0.32 &\cellcolor{yellow!30}0.01 &\cellcolor{yellow!30}70.91 &
              &\cellcolor{yellow!30}0.52 &\cellcolor{yellow!30}0.68 &\cellcolor{yellow!30}0.27 &\cellcolor{yellow!30}0.42 &\cellcolor{yellow!30}12.54 \\
        RoBERTa &\cellcolor{green!20}0.46 &\cellcolor{green!20}0.66 &\cellcolor{green!20}0.03 &\cellcolor{green!20}0.20 &\cellcolor{green!20}67.73 &
                ROBERTa-Syn &\cellcolor{green!20}0.51 &\cellcolor{green!20}0.67 &\cellcolor{green!20}0.45 &\cellcolor{green!20}0.62 &\cellcolor{green!20}\textbf{11.07} \\
              &\cellcolor{cyan!25}\textbf{0.90} &\cellcolor{cyan!25}\textbf{0.91} &\cellcolor{cyan!25}\textbf{0.54} &\cellcolor{cyan!25}\textbf{0.60} &\cellcolor{cyan!25}\textbf{28.14} &
              &\cellcolor{cyan!25}\textbf{0.99} &\cellcolor{cyan!25}\textbf{0.99} &\cellcolor{cyan!25}\textbf{0.87} &\cellcolor{cyan!25}\textbf{0.87} &\cellcolor{cyan!25}12.65 \\
        \hline
    \end{tabular}
    \end{threeparttable}
\end{table*}
\begin{table*}[ht]
    \centering
    \begin{threeparttable}
    \caption{Model Performance under Pruthi Attack Using Different Training Datasets.}
    \label{tab:evasion_attacks_three_datasets_3}
    \begin{tabular}{p{1cm}ccccc p{2cm}ccccc}
        \hline
        \multicolumn{6}{c}{\textbf{$D_{1}$} (Existing Phishing + Benign)} 
        & \multicolumn{6}{c}{\textbf{$D_{2}$} (Existing Phishing + PEEK phishing)} \\
        \hline
        Model & Acc & F1 & EVA-Acc & EVA-F1 & ASR\% & Model & Acc & F1 & EVA-Acc & EVA-F1 & ASR\% \\
        \hline
        & \cellcolor{yellow!30}0.64 & \cellcolor{yellow!30}0.71 & \cellcolor{yellow!30}0.54 & \cellcolor{yellow!30}0.62 & \cellcolor{yellow!30}15.36
        & & \cellcolor{yellow!30}0.64 & \cellcolor{yellow!30}0.71 & \cellcolor{yellow!30}0.54 & \cellcolor{yellow!30}0.62 & \cellcolor{yellow!30}13.32 \\

        SQ & \cellcolor{green!20}0.55 & \cellcolor{green!20}0.62 & \cellcolor{green!20}0.63 & \cellcolor{green!20}0.59 & \cellcolor{green!20}34.80
        & T5-Encoder & \cellcolor{green!20}0.55 & \cellcolor{green!20}0.60 & \cellcolor{green!20}0.47 & \cellcolor{green!20}0.38 & \cellcolor{green!20}43.28  \\

        & \cellcolor{cyan!25}\textbf{0.90} & \cellcolor{cyan!25}\textbf{0.91} & \cellcolor{cyan!25}\textbf{0.87} & \cellcolor{cyan!25}\textbf{0.88} & \cellcolor{cyan!25}\textbf{3.53}
        & & \cellcolor{cyan!25}\textbf{0.99} & \cellcolor{cyan!25}\textbf{0.99} & \cellcolor{cyan!25}\textbf{0.98} & \cellcolor{cyan!25}\textbf{0.99} & \cellcolor{cyan!25}\textbf{1}\\
        
        \hline

        & \cellcolor{yellow!30}0.59 & \cellcolor{yellow!30}0.74 & \cellcolor{yellow!30}0.57 & \cellcolor{yellow!30}0.73 & \cellcolor{yellow!30}\textbf{5.72}
        & & \cellcolor{yellow!30}0.52 & \cellcolor{yellow!30}0.68 & \cellcolor{yellow!30}0.52 & \cellcolor{yellow!30}0.68 & \cellcolor{yellow!30}11.07 \\

        ALBERT & \cellcolor{green!20}0.58 & \cellcolor{green!20}0.74 & \cellcolor{green!20}0.16 & \cellcolor{green!20}0.07 & \cellcolor{green!20}54.41
        & DistilBERT & \cellcolor{green!20}0.52 & \cellcolor{green!20}0.68 & \cellcolor{green!20}0.49 & \cellcolor{green!20}0.64 & \cellcolor{green!20}23.15 \\

        & \cellcolor{cyan!25}\textbf{0.88} & \cellcolor{cyan!25}\textbf{0.89} & \cellcolor{cyan!25}\textbf{0.78} & \cellcolor{cyan!25}\textbf{0.80} & \cellcolor{cyan!25}10.91
        & & \cellcolor{cyan!25}\textbf{0.99} & \cellcolor{cyan!25}\textbf{0.99} & \cellcolor{cyan!25}\textbf{0.98} & \cellcolor{cyan!25}\textbf{0.98} & \cellcolor{cyan!25}\textbf{0.81}\\
        
        \hline

        & \cellcolor{yellow!30}0.83 & \cellcolor{yellow!30}0.83 & \cellcolor{yellow!30}0.78 & \cellcolor{yellow!30}0.69 & \cellcolor{yellow!30}4.19
        & & \cellcolor{yellow!30}0.48 & \cellcolor{yellow!30}0.64 & \cellcolor{yellow!30}0.47 & \cellcolor{yellow!30}0.64 & \cellcolor{yellow!30}2.77 \\

        RoBERTa & \cellcolor{green!20}0.42 & \cellcolor{green!20}0.57 & \cellcolor{green!20}0.46 & \cellcolor{green!20}0.33 & \cellcolor{green!20}7.53
        & RoBERTa-Syn & \cellcolor{green!20}0.51 & \cellcolor{green!20}0.67 & \cellcolor{green!20}0.50 & \cellcolor{green!20}0.67 & \cellcolor{green!20}1.68 \\

        & \cellcolor{cyan!25}\textbf{0.91} & \cellcolor{cyan!25}\textbf{0.92} & \cellcolor{cyan!25}\textbf{0.87} & \cellcolor{cyan!25}\textbf{0.89} & \cellcolor{cyan!25}\textbf{1.23}
        & & \cellcolor{cyan!25}\textbf{0.99} & \cellcolor{cyan!25}\textbf{0.99} & \cellcolor{cyan!25}\textbf{0.99} & \cellcolor{cyan!25}\textbf{0.99} & \cellcolor{cyan!25}\textbf{0.2}\\

        \hline
    \end{tabular}
    \end{threeparttable}
\end{table*}

This section shows the performance of six detectors under implemented evasion attacks after training using three different datasets separately. Table \ref{tab:evasion_attacks_three_datasets_4}, and Tables \ref{tab:evasion_attacks_three_datasets_1} to \ref{tab:evasion_attacks_three_datasets_3} print out detection accuracy and F1-score in normal phishing attacks and evasion attacks under TextFooler, PWWS, Pruthi, and DeepwordBug Separately. Generally speaking, for defending against adversarial attacks, detectors are trained by PEEK phishing to achieve the best performance, albeit with the performance of several models that are trained using non-PEEK phishing is better than that of PEEK phishing. Further PEEK against perturbative attacks are shown in Appendix~\ref{appendix:evasion_attacks_other_results}, which adds a column of attack success rate to further indicate the obvious increase in evasion detection accuracy. The datasets provided by PEEK mitigate the attacks steadily and consistently, which is crucial for detectors.

\subsection{PEEK phishing Defend Against Other Evasion Attacks}\label{appendix:evasion_attacks_other_results}
As discussed in the main text, the inclusion of PEEK phishing in the training dataset significantly enhances the robustness of phishing detectors, particularly in maintaining high detection accuracy when subjected to adversarial attacks. This improvement is evident across multiple evaluation metrics, including accuracy, F1-score, and adversarial evaluation metrics (EVA-Acc and EVA-F1), results shown in Tables \ref{tab:merged_performance_1}, \ref{tab:merged_performance_2}, and \ref{tab:merged_performance_3}. This section shows model performance under TextFooler, Pruthi, and PWWS before and after training with generated content. Overall, the generated content improves model performance significantly. 

However, different models exhibit varying degrees of robustness when exposed to different types of adversarial attacks. For instance, while RoBERTa-Syn consistently achieves high scores across multiple metrics, other models, such as T5-Encoder and DistilBERT, show more variable performance, suggesting that their ability to withstand adversarial perturbations depends on the nature of the attack. These observations highlight the importance of model selection in adversarial phishing detection. Aligning model choice with specific threat types encountered in real-world scenarios can improve detection reliability. Moreover, training with adversarially augmented datasets, such as those incorporating PEEK phishing, offers a promising strategy for enhancing model robustness.

\begin{table}[H]
    \centering
    \begin{threeparttable}
    \caption{Model Performance under TextFooler Attack Before and After PEEK phishing Fine-Tuning.}
    \label{tab:merged_performance_1}
    \begin{tabular}{lcccccc}
        \hline
        Status & Model & Acc & F1 & EVA-Acc\tnote{1} & EVA-F1\tnote{2} & ASR\%\tnote{3} \\
        \hline
        \multicolumn{7}{l}{\textit{\textbf{$D_{1}$} (Existing Phishing + Benign)}} \\
        \hline
         & SQ & 0.32 & 0.42 & 0.03 & 0.01 & 90.18 \\
        Before & ALBERT & 0.54 & 0.57 & 0.02 & 0.02 & 95.99 \\
         & RoBERTa & 0.54 & 0.57 & 0.3 & 0.07 & 44.44 \\
        \hline
        \multicolumn{7}{l}{\textit{\textbf{$D_{2}$} (Existing Phishing + PEEK phishing)}} \\
        \hline
         & T5-Encoder & 0.53 & 0.68 & 0.19 & 0.32 & 34.2 \\
        Before & DistilBERT & 0.42 & 0.51 & 0.15 & 0.12 & 33.7 \\
         & RoBERTa-Syn & 0.44 & 0.35 & 0.13 & 0.08 & 31.2 \\
        \hline
        \multicolumn{7}{c}{\textbf{--- Separation Line: Before and After ---}} \\
        \hline
        \multicolumn{7}{l}{\textit{\textbf{$D_{1}$} (Existing Phishing + Benign)}} \\
        \hline
         & SQ & 0.88 & 0.89 & 0.88 & 0.77 & 19.6 \\
        After & ALBERT & 0.88 & 0.89 & 0.55 & 0.57 & 49.55 \\
         & RoBERTa & 0.90 & 0.91 & 0.72 & 0.74 & 19.6 \\
        \hline
        \multicolumn{7}{l}{\textit{\textbf{$D_{2}$} (Existing Phishing + PEEK phishing)}} \\
        \hline
         & T5-Encoder & 0.99 & 0.99 & 0.89 & 0.90 & 3.11 \\
        After & DistilBERT & 0.99 & 0.99 & 0.87 & 0.88 & 12.68 \\
         & RoBERTa-Syn & 0.99 & 0.99 & 0.94 & 0.94 & 5.42 \\
        \hline
    \end{tabular}
    \begin{tablenotes}
        \item[1] \textit{EVA-Acc}: Accuracy under evasion attacks.
        \item[2] \textit{EVA-F1}: F1-Score under evasion attacks.
        \item[3] \textit{ASR}: Attack Success Rate, indicating the percentage of data bypassing detectors.
    \end{tablenotes}
    \end{threeparttable}
\end{table}
\begin{table}[H]
    \centering
    \caption{Model Performance under Pruthi Attack Before and After PEEK phishing Fine-Tuning. 
    }
    \label{tab:merged_performance_2}
    
    \begin{tabular}{lcccccc}
        \hline
        Status & Model & Acc & F1 & EVA-Acc & EVA-F1 & ASR\% \\
        \hline
        \multicolumn{7}{l}{\textit{\textbf{$D_{1}$} (Existing Phishing + Benign)}} \\
        \hline
         & SQ & 0.32 & 0.42 & 0.26 & 0.31 & 19.02 \\
        Before & ALBERT & 0.54 & 0.57 & 0.26 & 0.34 & 51.82 \\
         & RoBERTa & 0.54 & 0.57 & 0.52 & 0.34 & 2.96 \\
        \hline
        \multicolumn{7}{l}{\textit{\textbf{$D_{2}$} (Existing Phishing + PEEK phishing)}} \\
        \hline
         & T5-Encoder & 0.53 & 0.68 & 0.49 & 0.66 & 4.4 \\
        Before & DistilBERT & 0.42 & 0.51 & 0.27 & 0.26 & 18.97 \\
         & RoBERTa-Syn & 0.44 & 0.35 & 0.33 & 0.21 & 11.2 \\
        \hline
        \multicolumn{7}{c}{\textbf{--- Separation Line: Before and After ---}} \\
        \hline
        \multicolumn{7}{l}{\textit{\textbf{$D_{1}$} (Existing Phishing + Benign)}} \\
        \hline
         & SQ & 0.90 & 0.91 & 0.87 & 0.88 & 3.53 \\
        After & ALBERT & 0.88 & 0.89 & 0.78 & 0.80 & 10.91 \\
         & RoBERTa & 0.91 & 0.92 & 0.87 & 0.89 & 1.23 \\
        \hline
        \multicolumn{7}{l}{\textit{\textbf{$D_{2}$} (Existing Phishing + PEEK phishing)}} \\
        \hline
         & T5-Encoder & 0.99 & 0.99 & 0.98 & 0.99 & 1 \\
        After & DistilBERT & 0.99 & 0.99 & 0.98 & 0.98 & 0.81 \\
         & RoBERTa-Syn & 0.99 & 0.99 & 0.99 & 0.99 & 0.2 \\
        \hline
    \end{tabular}
\end{table}
\begin{table}[H]
    \centering
    \caption{Model Performance under PWWS Attack Before and After PEEK phishing Fine-Tuning. 
    }
    \label{tab:merged_performance_3}
    
    \begin{tabular}{lcccccc}
        \hline
        Status & Model & Acc & F1 & EVA-Acc & EVA-F1 & ASR\% \\
        \hline
        \multicolumn{7}{l}{\textit{\textbf{$D_{1}$} (Existing Phishing + Benign)}} \\
        \hline
         & SQ & 0.32 & 0.42 & 0.08 & 0.08 & 74.23 \\
        Before & ALBERT & 0.54 & 0.57 & 0.08 & 0.06 & 85.4 \\
         & RoBERTa & 0.54 & 0.57 & 0.37 & 0.24 & 31.11 \\
        \hline
        \multicolumn{7}{l}{\textit{\textbf{$D_{2}$} (Existing Phishing + PEEK phishing)}} \\
        \hline
         & T5-Encoder & 0.53 & 0.68 & 0.26 & 0.42 & 27.2 \\
        Before & DistilBERT & 0.42 & 0.51 & 0.15 & 0.12 & 33.4 \\
         & RoBERTa-Syn & 0.44 & 0.35 & 0.16 & 0.09 & 28.6 \\
        \hline
        \multicolumn{7}{c}{\textbf{--- Separation Line: Before and After ---}} \\
        \hline
        \multicolumn{7}{l}{\textit{\textbf{$D_{1}$} (Existing Phishing + Benign)}} \\
        \hline
         & SQ & 0.88 & 0.89 & 0.55 & 0.52 & 38.98 \\
        After & ALBERT & 0.88 & 0.89 & 0.31 & 0.41 & 65.16 \\
         & RoBERTa & 0.90 & 0.91 & 0.55 & 0.60 & 28.14 \\
        \hline
        \multicolumn{7}{l}{\textit{\textbf{$D_{2}$} (Existing Phishing + PEEK phishing)}} \\
        \hline
         & T5-Encoder & 0.99 & 0.99 & 0.88 & 0.88 & 11.45 \\
        After & DistilBERT & 0.99 & 0.99 & 0.87 & 0.88 & 12.27 \\
         & RoBERTa-Syn & 0.99 & 0.99 & 0.87 & 0.87 & 12.65 \\
        \hline
    \end{tabular}
\end{table}

\subsection{Evasion Model Parameters}\label{appendix:EVA_model_parameters}
Here, we provide the training parameters for PEEK phishing models designed to defend against different types of attacks. We implemented two categories of phishing detectors. The first category, including Detectors ALBERT, RoBERTa, SQ~\cite{mehdi2023adversarial}, are trained to distinguish between phishing emails and general emails. The second category, comprising Detectors DistilBERT, RoBERTa-LLMs \cite{jamal2024improved} and T5-Encoder \cite{bethany2024large} aimed at differentiating between handwritten phishing emails and machine-generated phishing emails. 

This division allows for a more targeted defense strategy based on the nature of the phishing content. Such categorization also facilitates evaluating model robustness under different phishing generation paradigms. The performance of each model varies depending on the type of attack encountered. The specific models and their training parameters are summarized in Table \ref{tab:evasion_attack_model_parameters_1} and \ref{tab:evasion_attack_model_parameters_2}.
\begin{table}[H]
\centering
\caption{Fine-Tuning Parameters of $Detector_{1}$. For diverse attacks and detectors, parameters are also different, especially the learning rate. The following are the training parameters of the detector to defend the Evasion Attacks using PEEK phishing.}
\label{tab:evasion_attack_model_parameters_1}
\begin{tabular}{llp{1.2cm}p{0.8cm}c}
\hline
Models & Attacks & Learning Rate & Batch Size & Epochs \\
\hline
SQ & TextFooler & 5e-05 & \multirow{4}{*}{\centering 16} & \multirow{4}{*}{\centering 10} \\
   & PWWS       & 5e-05 &                                   &                                \\
   & Pruthi     & 3e-05 &                                   &                                \\
   & DeepWordBug& 1e-05 &                                   &                                \\
\hline
ALBERT & TextFooler & 2e-05 & \multirow{4}{*}{\centering 16} & \multirow{4}{*}{\centering 10} \\
       & PWWS       & 2e-05 &                                &                                \\
       & Pruthi     & 2e-05 &                                &                                \\
       & DeepWordBug& 2e-05 &                                &                                \\
\hline
RoBERTa & TextFooler & 5e-05 & \multirow{4}{*}{\centering 16} & \multirow{4}{*}{\centering 10} \\
        & PWWS       & 5e-05 &                                &                                \\
        & Pruthi     & 2e-05 &                                &                                \\
        & DeepWordBug& 2e-05 &                                &                                \\
\hline
\end{tabular}
\end{table}

\begin{table}[H]
\centering
\caption{Fine-Tuning Parameters of $Detector_{2}$. For diverse attacks and detectors, parameters are also different, especially the learning rate. The following are the training parameters of the detector to defend against Evasion Attacks using PEEK phishing.}
\label{tab:evasion_attack_model_parameters_2}
\begin{tabular}{llp{1.2cm}p{0.8cm}c}
\hline
Models & Attacks & Learning Rate & Batch Size & Epochs \\
\hline
T5-Encoder & TextFooler & 3e-05 & \multirow{4}{*}{\centering 16} & \multirow{4}{*}{\centering 10} \\
           & PWWS       & 2e-05 &                                &                                \\
           & Pruthi     & 5e-05 &                                &                                \\
           & DeepWordBug& 5e-05 &                                &                                \\
\hline
DistilBERT & TextFooler & 2e-05 & \multirow{4}{*}{\centering 16} & \multirow{4}{*}{\centering 10} \\
           & PWWS       & 2e-05 &                                &                                \\
           & Pruthi     & 1e-05 &                                &                                \\
           & DeepWordBug& 3e-05 &                                &                                \\
\hline
RoBERTa-Syn & TextFooler & 3e-05 & \multirow{4}{*}{\centering 16} & \multirow{4}{*}{\centering 10} \\
            & PWWS       & 5e-05 &                                &                                \\
            & Pruthi     & 5e-05 &                                &                                \\
            & DeepWordBug& 5e-05 &                                &                                \\
\hline
\end{tabular}
\end{table}

\section{Persuasion Principle Usage Differences among Diverse Phishing Dominants.}\label{appendix:extra_principles_frequency_rest}
\begin{figure}[H]
\centering
\includegraphics[width=1\linewidth, clip, trim = 0 0 0 0.9cm]{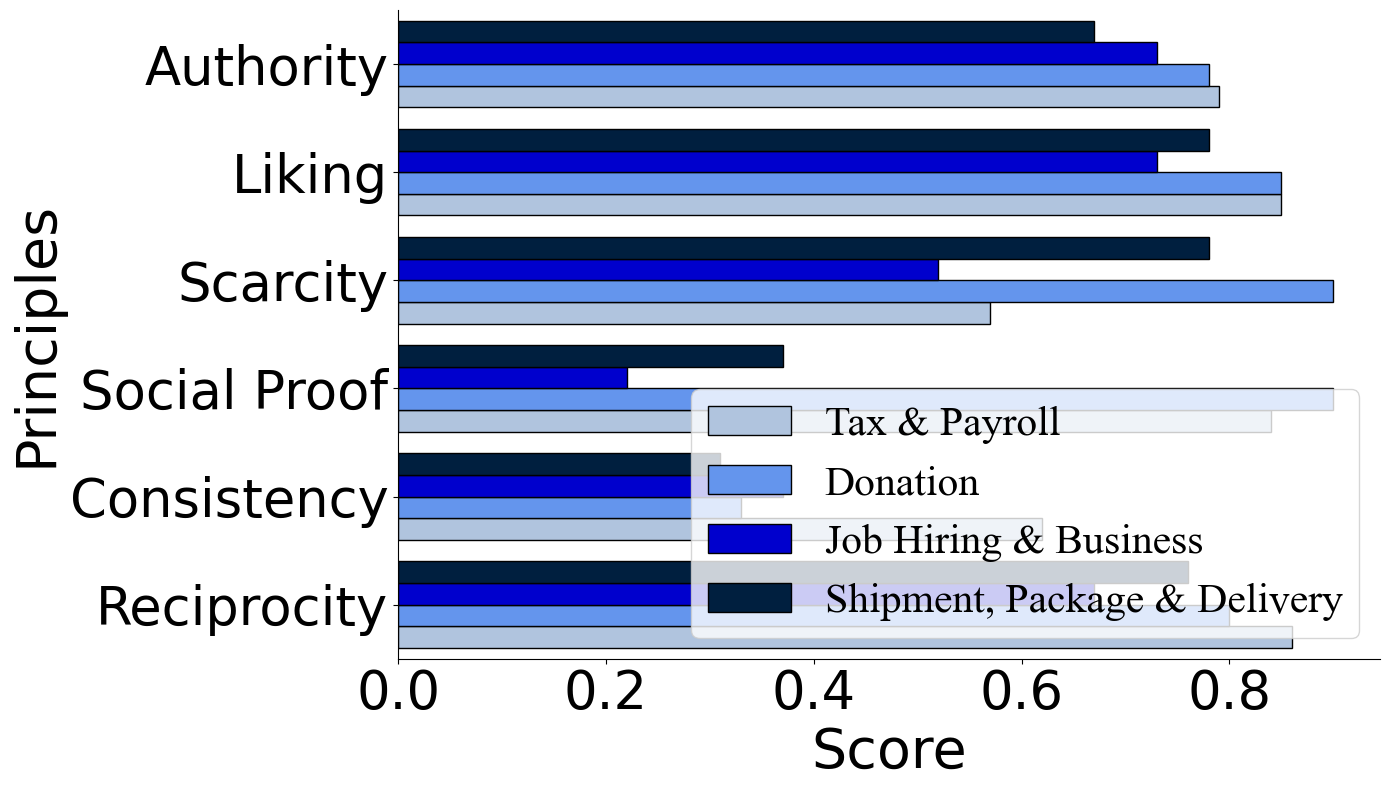}
\caption{Persuasion Principle Usage Across Phishing Themes. Higher scores indicate greater use of persuasive principles. Different domains exhibit distinct deceptive tactics.}
\label{fig_extra_principles_frequency_rest}

\end{figure}

Fig.~\ref{fig_extra_principles_frequency_rest} presents the distribution of persuasion principles utilized in phishing emails categorized under different themes, including banking, donations, job hiring, and shipment/package delivery. The results indicate that phishing attacks employ distinct psychological manipulation strategies depending on the thematic context of the email. For instance, certain principles, such as \textit{Authority} and \textit{Social Proof}, may be more prevalent in banking-related phishing attempts, whereas \textit{Reciprocity} and \textit{Scarcity} might be more frequently observed in donation scams. Understanding these variations in attack strategies across different phishing themes is crucial, as it enables researchers and security practitioners to develop more effective countermeasures. Additionally, raising awareness among users about these persuasive tactics can help mitigate the risk of falling victim to phishing attempts, ultimately reducing potential financial and informational losses.

\vspace{8cm}

\section{Keywords of Across Topic Phishing Emails}\label{appendix:expand_education_topic_keywords}
This section discusses the expansion of phishing email topics, with themes and corresponding keywords generated using ChatGPT. A total of seven distinct main topics are covered, each containing five subtopics. The corresponding keywords are presented in the third column of Table \ref{tab:Extra_Topic_Keywords}.

\begin{table}[ht]
    \centering
    \caption{Topic keywords of Expanded Dominant Phishing Emails.}
    \resizebox{\linewidth}{!}{
    \begin{tabular}{p{1cm}lp{8cm}}
    \hline
       \textbf{Dominant} &\textbf{Topics} &\textbf{T-Keywords}  \\
    \hline
        & Online Learning & online, course, student, platform, access, education, flexible, virtual, materials, instructor \\
        & Education Technology  & technology, tools, integration, innovation, resources, digital, learning, platforms, software, support \\
    Education& Special Education & disabilities, support, needs, services, inclusion, learning, strategies, resources, intervention, educators\\
       & Teacher Training  & professional, development, training, educators, skills, techniques, pedagogy, teaching, workshops, strategies  \\
       & Early Childhood Education & development, learning, activities, play, curriculum, teaching, preschool, skills, educators, engagement \\
    \hline
    & Company Review& performance, feedback, culture, growth, management, innovation, strategy, vision, communication, leadership\\
    & Phone Message& Call, voicemail, greeting, notification, recording, message, audio, tone, playback, timestamp\\
   Audio& Delivery&Delivery, notification, package, update, courier, logistics, arrival, shipped, tracking, dispatch \\
    Message & Speech Analytics &Transcription, Emotion, Sentiment, Keyword, Tone, Compliance, Call, Speaker, Accuracy, Detection \\
    & Voice Recognition&Recognition, Authentication, Biometric, Signal, Frequency, Speech, Detection, Acoustic, Neural, Accuracy \\
    \hline
    &Data Analytics & insights, reporting, statistics, visualization, trends, analytics, performance, KPIs, dashboard, data\\
    & File Management& upload, download, folder, storage, organize, file,metadata, sync,archive, access\\
  Document & Security and Privacy& security, permission, authentication, encryption,access, confidential,user, policy, audit, protocol\\
    Shared& User Experience&interface, navigation, user-friendly, feedback,support, performance,design, accessibility, tutorials, customization \\
    & Workflow and Precess &approval, process, task, deadline, assignment, review, status, tracking, integration, automation \\
    \hline
    & Impact of Donations& impact, change, community, funds, beneficiaries, support, projects, programs, success, contributions \\
    & Corporate Giving&  corporate, donations, partnership, social, responsibility, community, impact, engagement, support, initiatives\\
   Donation & Online Donations &  website, donate, online, platform, contribution, secure, payment, fundraising, transaction, charity\\
    & Volunteering & volunteer, service, community, help, support, outreach, organization, event, engagement, contribution \\
    & Philanthropy& philanthropy, wealthy, donors, giving, initiatives, impact, organizations, community, support, engagement \\
    \hline
    &Job Posting &advertisement, vacancy, description, requirements, salary, benefits, location, company, deadlines, applicants \\
    &Networking&connections, professionals, industry, events, relationships, partnerships, collaborations, referrals, social, engagement \\
   Job & Requirement & hiring, candidates, interview, job, application, employer, talent, position, skills, qualifications\\
    Hiring &Work Environment & culture, values, team, collaboration, flexibility, atmosphere, support, communication, workplace, diversity\\
    & Performance Evaluation & reviews, feedback, goals, metrics, productivity, outcomes, assessments, criteria, improvement, appraisals\\
    \hline
    & Delivery Process&shipment, tracking, confirmation, carrier, logistics, delivery, package, route, customer, service \\
    & Shipment Methods& air, ground, express, freight, courier, international, domestic, parcel, mail, shipment\\
   Shipment & Package Handling & fragile, secure, packaging, labeling, sorting, warehouse, inspection, delivery, transportation, inventory\\
    & E-Commerce Shipping&online, order, checkout, shipping, delivery, confirmation, address, package, tracking, customer \\
    & International Shipping& customs, duties, tariffs, regulations, documents, international, delivery, freight, zone, clearance\\
    \hline
    & Deductions and Credits& Deductions, Credits, Taxable, Income, Expenses, Eligibility, Claiming, Adjustments, Taxpayers, Benefits\\
    & Employment Taxes& Employment, Taxes, FICA, Medicare, Unemployment, Withholding, Liabilities, Contributions, Employer, Employee\\
  Tax/Payroll & Tax Forms & Forms, 1040, 1099, W-2, Instructions, Filing, Deadlines, Information, Compliance, Reporting\\
    & Payroll Processing&Payroll, Processing, Employees, Payments, Deductions, Calculations, Timecards, Reports, Software, Accuracy \\
    & Wage Reporting&Wages, Reporting, Forms, IRS, 1099, 1040, Year-End, Employees, Withholdings, Compliance \\
    \hline
    \end{tabular}
    }
    \label{tab:Extra_Topic_Keywords}
\end{table}
\end{appendices}

\end{document}